\documentclass[11pt,draftcls,onecolumn]{IEEEtran}
\usepackage{booktabs}
\usepackage{amsfonts,bm,amsmath,comment,color,cite,amssymb,subfigure,ntheorem}
\usepackage{array}
\usepackage[]{graphicx}
\usepackage[linesnumbered,ruled]{algorithm2e}
\usepackage{multirow}
\usepackage[colorlinks,urlcolor=blue,linkcolor=blue,citecolor=blue]{hyperref}
\usepackage{makecell}

\newtheorem{Prop}{Proposition}
\theoremnumbering {arabic}
\usepackage{algpseudocode}
\usepackage{color}
\usepackage{float}
\usepackage[utf8]{inputenc}

\makeatletter

\newcommand{\Rmnum}[1]{\expandafter\@slowromancap\romannumeral #1@}
\makeatother

\def\ba{{\mathbf a}}
\def\bb{{\mathbf b}}
\def\bc{{\mathbf c}}

\def\be{{\mathbf e}}
\def\bg{{\mathbf g}}
\def\bh{{\mathbf h}}

\def\bn{{\mathbf n}}

\def\bs{{\mathbf s}}

\def\bu{{\mathbf u}}

\def\bw{{\mathbf w}}
\def\bx{{\mathbf x}}
\def\by{{\mathbf y}}
\def\bz{{\mathbf z}}
\def\bA{{\mathbf A}}
\def\bB{{\mathbf B}}
\def\bC{{\mathbf C}}
\def\bD{{\mathbf D}}

\def\bF{{\mathbf F}}

\def\bH{{\mathbf H}}
\def\bI{{\mathbf I}}

\def\bM{{\mathbf M}}
\def\bN{{\mathbf N}}

\def\bQ{{\mathbf Q}}
\def\bR{{\mathbf R}}
\def\bS{{\mathbf S}}
\def\bT{{\mathbf T}}

\def\bW{{\mathbf W}}
\def\bX{{\mathbf X}}
\def\bY{{\mathbf Y}}
\def\bZ{{\mathbf Z}}

\def\blambda{\boldsymbol{\lambda}}

\def\bgamma{{\pmb {\gamma}}}
\def\complexC{{\mathbb{C}}}
\def\realR{{\mathbb{R}}}
\def\expE{{\mathbb{E}}}

\def\bzero{{\mathbf{0}}}
\def\u{{\bm{\upsilon}}}
\def\v{{\bm{\nu}}}
\def\L{{\mathcal{L}}}
\def\O{{\mathcal{O}}}

\begin{document}

\title{Dual-Function Beamforming Design For Multi-Target Localization and Reliable Communications}
\author{Bo~Tang,~
             Da~Li,
             Wenjun~Wu,
             Astha~Saini,
             Prabhu~Babu,
             and~Petre~Stoica,~\IEEEmembership{Life Fellow,~IEEE}
\thanks{}
\thanks{The work of Bo Tang was supported in part by the National Natural Science Foundation of China under Grant 62171450, in part by the Anhui Provincial Natural Science Foundation under Grant 2108085J30, and in part by the Research Plan of National University of Defense Technology under Grant 23-ZZCX-JDZ-42. The work of Petre Stoica was supported by the Swedish Research Council under Grants 2017-04610, 2016-06079, and 2021-05022 (\it{Corresponding author: Bo Tang}\textrm).}
\thanks{Bo Tang, Da Li, and Wenjun Wu are with the College of Electronic Engineering, National University of Defense Technology, Hefei 230037, China  (email: tangbo06@gmail.com).}
\thanks{Astha~Saini and Prabhu~Babu are with the Centre for Applied Research in Electronics, Indian
Institute of Technology, Delhi 110016, India (email: Prabhu.Babu@care.iitd.ac.in).}
\thanks{Petre Stoica is with the Department of Information Technology, Uppsala University, 75105 Uppsala, Sweden (email: ps@it.uu.se).}
} \markboth{Preprint}
{Shell \MakeLowercase{\textit{et al.}}: A Sample Article Using IEEEtran.cls for IEEE Journals}

\maketitle

\begin{abstract}
This paper investigates the transmit beamforming design for multiple-input multiple-output systems to support both multi-target localization and multi-user communications. To enhance the target localization performance, we derive the asymptotic Cram\'{e}r-Rao bound (CRB) for target angle estimation by assuming that the receive array is linear and uniform. Then we formulate a beamforming  design problem based on minimizing an upper bound on the asymptotic CRB (which is shown to be equivalent to {maximizing} the harmonic mean of the weighted beampattern responses at the target directions). Moreover, we impose a constraint on the SINR of each received communication signal to guarantee reliable communication performance. Two iterative algorithms are derived to tackle the non-convex design problem: one is based on the alternating direction method of multipliers, and the other uses the  majorization-minimization technique to solve an equivalent minimax problem. Numerical results show that, through elaborate dual-function beamforming matrix design, the proposed algorithms can simultaneously achieve superior angle estimation performance as well as high-quality multi-user communications.
\end{abstract}

\begin{IEEEkeywords}
MIMO systems, dual-function radar and communications (DFRC), beamforming design, multi-target localization, angle estimation, CRB, multi-user communications.
\end{IEEEkeywords}

\section{Introduction}\label{sec:Introduction}
With the development of next-generation wireless communications and the Internet of Things, the proliferation of radio frequency equipment has led to a growing demand for access to the spectrum. At the same time, to finely extract the features and identify the targets, the range resolvability of radar systems is continuously evolving. To achieve high range resolution, radar systems should possess sufficiently large bandwidth. However, the scarcity of the spectral resources will inevitably result in conflicts between radar and wireless communication systems. As a consequence, the mutual interference between them will degrade the performance of both systems \cite{Challenge}.
To reduce the mutual interference and efficiently utilize the spectral resources, a variety of solutions have been proposed, including designing spectrally constrained waveforms \cite{Rowe2014shape,Aubry2016Optimization,Tang2019Spectrally,Yang2022Multispectral,Li2023Multispectral}, and collaborative design of radar and communication systems \cite{Li2016codesign,Qian2018Codesign,Cheng2019CoDesign,Qian2021Coexistence,Li2023CoDesign}.
Another highly promising strategy to improve the spectral efficiency is the development of dual-function radar-communication (DFRC) systems (also called joint radar and communication systems, or integrated sensing and communication systems) \cite{Hassanien2019SPM,Zheng2019Coexistence,Liu2023SeventyYears}.
Based on shared antenna arrays and hardware components, DFRC systems can support both radar sensing and data communications via transmitting integrated waveforms. Compared with the other solutions, the DFRC systems reduce the number of antennas, the cost, the power consumption as well as the size. Due to these advantages, the development of DFRC systems has attracted considerable interest from both academia and industry \cite{Hassanien2019SPM,Zheng2019Coexistence,Liu2023SeventyYears,tavik2005advanced}. 

Note that DFRC systems based on conventional arrays have difficulties achieving simultaneous sensing and multi-user communications. 
Unlike traditional phased arrays, multiple-input multiple-output (MIMO) arrays have the capability of transmitting diverse waveforms through different antennas \cite{MIMO,tse2005fundamentals}. By utilizing the waveform diversity, MIMO systems not only have better detection, estimation, and communication performance \cite{telatar1999capacity,Paulraj2004mimo,Li2007Identifiability,Chen2008MIMOSTAP,Tang2020TSP}, but also have the potential to support multiple functions \cite{Hassanien2016InformationEmbedding,LiuFan2018DFRC,Liu2020Joint,Tang2020DFRC,Yu2022TSP,
Tang2022TSP,Tsinos2021DFRC, Wu2023DFRC,Wen2023Transceiver, Wang2023IEEESJ,LI2024104375,LIU2022CRB,Guo2023Bistatic}.
Therefore, there are considerable efforts to design DFRC systems based on  MIMO arrays (which are also called MIMO DFRC systems).
A key problem in MIMO DFRC systems is the design of transmit waveforms or transmit beamforming matrix.
In \cite{Hassanien2016InformationEmbedding,LiuFan2018DFRC,Liu2020Joint,Tang2020DFRC,Yu2022TSP}, the authors considered the design of dual-function waveforms or beamforming matrix, whose purpose was to approximate a desired beampattern (for sensing) and support communications. The main differences between these works lie in how they deliver the information bits. The delivery methods include controlling the sidelobes of the beampattern, minimizing the multi-user interference (MUI), and varying the spectral shape of the transmit signals.
In \cite{Tsinos2021DFRC,Wu2023DFRC,Wen2023Transceiver}, the authors addressed the waveform design problem for MIMO DFRC systems in the presence of clutter. To suppress the clutter and improve the signal-to-interference-plus-noise-ratio (SINR), joint design of transmit waveforms and receive filters was proposed. Moreover, communication related constraints (e.g., SINR constraint on the received communication signals or MUI constraint) were enforced to support multi-user communications.
In \cite{Wang2023IEEESJ,LI2024104375}, information-theoretic approaches were investigated to design MIMO DFRC systems, where the authors aimed to maximize the relative entropy between the probability density function of the observations under two hypotheses as well as minimize the MUI.

In this paper, we consider the transmit beamforming design for MIMO DFRC systems. The design objective is to enhance the multi-target localization performance and guarantee the multi-user communication quality of service (QOS). To this end, we consider the minimization of  the Cram\'{e}r-Rao bound (CRB) for target angle estimation. Note that the CRB considered in this paper is different from that in {\cite[Section II.C]{LIU2022CRB}}, \cite{Hua2023Tradeoff}, where the CRB for  the target response matrix estimation (which has many more unknowns than the target angles) is used as the design metric. Moreover, the design objective proposed in this paper does not require that the waveform covariance matrix is invertible. As a result, the data stream augmentation proposed in \cite{LIU2022CRB}, which results in energy waste and additional interference in the received communication signals, becomes unnecessary. To make the optimization problem tractable, we derive an upper bound on the asymptotic CRB and use the upper bound as the design metric. Moreover, we impose an SINR constraint on each received communication signals to guarantee the communication QOS. To solve the beamforming design problem, two algorithms are derived. One algorithm is based on the alternating direction method of multipliers (ADMM). The other algorithm resorts to a variational form of the objective and transforms the design problem into a minimax problem.  Then a \textbf{m}ajorization-\textbf{m}inimization based approach is developed \textbf{for} the \textbf{m}ini\textbf{m}ax problem (we call this algorithm MM$4$MM for short). Numerical examples are provided to show the performance of the proposed algorithms.

The rest of this paper is organized as follows.
In Section \ref{sec:Signal Model and Problem Formulation}, the signal models are established.
Then, the CRB for target angle estimation is analyzed and the beamforming design problem is formulated.
In Section \ref{Sec:Design}, two iterative algorithms are derived to tackle this problem.
In Section \ref{Sec:Numerical Results}, numerical examples are provided to illustrate the performance of the proposed design algorithms.
Finally, we conclude the paper in Section \ref{Sec:Conclusion}.

\textit{Notations}:
$\bA$, $\ba$, and $a$ stands for matrices, vectors, and scalars, respectively.
$\realR$ and $\complexC$ represent the domain of real-valued and complex-valued numbers.
$\bI$ denotes the identity matrix with the size determined by the subscript.
$(\cdot)^\top$,  $(\cdot)^*$, and $(\cdot)^\dagger$ indicate the transpose, conjugate,  and conjugate transpose.
$\textrm{Diag}(\bx)$ is the diagonal matrix with the diagonal elements being $\bx$.
$\textrm{BlkDiag}(\bA;\bB)$ denotes the block diagonal matrix formed by $\bA$ and $\bB$.
$\textrm{tr}(\cdot)$ represents the trace of a matrix. $\left\| \cdot \right\|_{1}$, $\left\| \cdot \right\|_{2}$, and $\left\| \cdot \right\|_\textrm{F}$ denote the $\ell_1$ norm, the Euclidian norm, and the Frobenius norm.
$\textrm{Re}(\cdot)$, $\textrm{Im}(\cdot)$, and $\arg(\cdot)$ indicate the real part, the imaginary part, and the argument of a complex-valued scalar/vector/matrix. $\odot$ stands for the Hadamard (element-wise) matrix product.  The letter $\textrm{j}$ denotes the imaginary unit (i.e., $\textrm{j} = \sqrt{-1}$). $\bA\succ \bm{0}$ ($\bA\succeq \bm{0}$) indicates that $\bA$ is positive definite (semi-definite). $\textrm{unvec}_{m,n}(\bx)$ denotes the operation of arranging $\bx_{mn\times 1}$ columnwise into $\bX_{m\times n}$. $\lambda_{\min}(\bA)$ ($\lambda_{\max}(\bA)$) represents the smallest (largest) eigenvalue of ${\bA}$.
Finally, $\expE\{\cdot \}$ denotes the expectation of a random variable.
\section{Signal Model and Problem Formulation}\label{sec:Signal Model and Problem Formulation}
\begin{figure}[!htbp]
	\setlength{\abovecaptionskip}{0.cm}
	\setlength{\abovecaptionskip}{0.cm}
	\setlength{\belowdisplayskip}{0pt}
	\centering
	\includegraphics[width= 0.45 \textwidth] {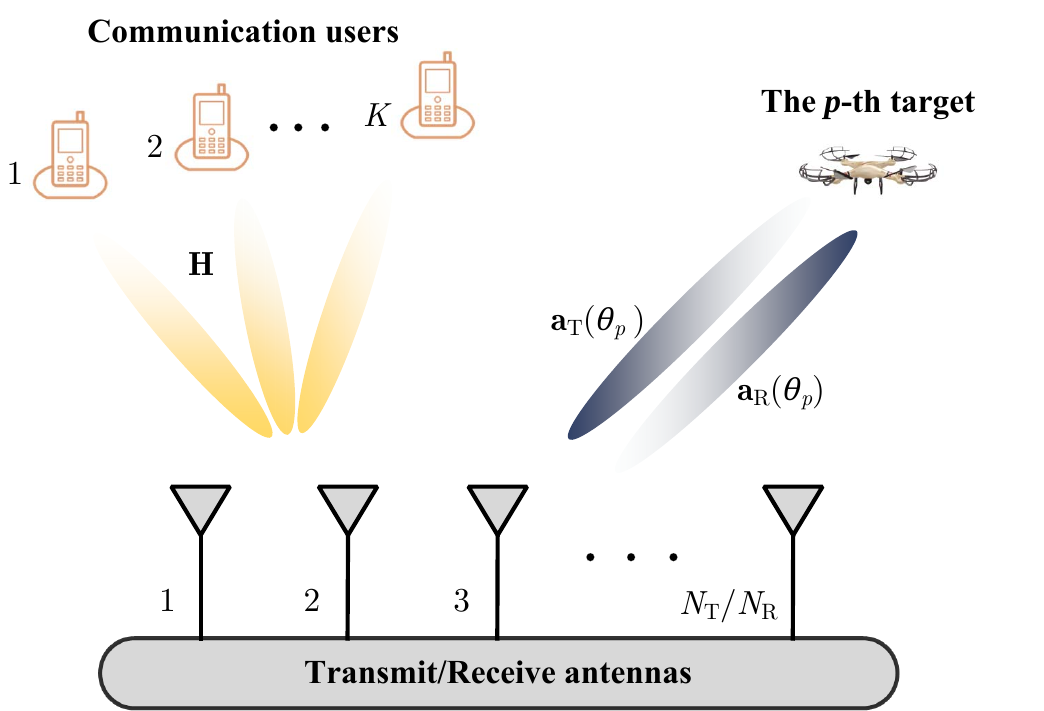}
	\caption{Illustration of the MIMO DFRC system.}
	\label{fig_1}
\end{figure}
\subsection{Communication Model}
We consider a MIMO DFRC system with $N_{\textrm{T}}$ transmit antennas and $N_{\textrm{R}}$ receive antennas, as shown in \figurename~\ref{fig_1}. Assume that the transmit and the receive antenna arrays are linear. Moreover, the inter-element spacing between the receive array elements is equal to half wavelength (i.e., the receive array is a standard uniform linear array (ULA)).
Let $K (K \le N_{\textrm{T}})$ denote the number of downlink communication users served by the DFRC system, and $\bs_k\in \complexC^{L \times 1}$ denote the data stream transmitted toward the $k$th user, $k=1,2,\cdots,K$, where $L$ is the code length. To reduce the multi-user interference in the communication signals, we consider transmit beamforming. The transmit waveform matrix after beamforming can be written as
\begin{align}\label{eq:X}
	\bX = \bW \bS = \sum_{k=1}^K {\bw_k\bs_k^\top} {, }
\end{align}
where $\bW = [\bw_{1}, \bw_{2}, \cdots ,\bw_{K}] \in \complexC^{N_{\textrm{T}} \times K} $ is the beamforming matrix, and $\bS = [\bs_{1}, \bs_{2}, \cdots ,\bs_{K}]^{\top} \in \complexC^{K \times L}$ is the data matrix.
Assume that the $K$ data streams are independent. Moreover, each data stream has an average power of 1. Thus, we have
\begin{align}\label{eq:S}
	\frac{1}{L} \bS \bS^{\dagger} \approx \bI_{K}.
\end{align}

The signal received by the $k$th user can be written as
\begin{align}\label{eq:k-th user}
	\by_{\textrm{C},k} = \bh_{k}^{\dagger} \bX + \bn_{\textrm{C}, k} = \bh_{k}^{\dagger}\bW\bS  + \bn_{\textrm{C},k},
\end{align}
where $\bh_k\in \complexC^{N_{\textrm{T}} \times 1}$ denotes the channel response vector between the DFRC system and the $k$th communication user, and $\bn_{\textrm{C},k}$ is the receiver noise, $k = 1, 2,  \cdots,K$. Thus,  by using \eqref{eq:X}, the SINR of the $k$th user can be defined as \cite{Liu2020Joint,Zhao2022ISAC,LIU2022CRB}
\begin{align}\label{eq:k-th user SINR}
	\rho_{k}
 =& \dfrac{|\bh_{k}^{\dagger}\bw_{k}|^{2}\mathbb{E}\{\|\bs_k\|_2^2\}}{\sum_{\breve{k} = 1 , \breve{k} \neq k}^{K} |\bh_{k}^{\dagger}\bw_{\breve{k}}|^{2}\mathbb{E}\{\|\bs_{\breve{k}}\|_2^2\} + \mathbb{E}\{\|\bn_{\textrm{C},k}\|_2^2\}  }\nonumber\\
=&\dfrac{|\bh_{k}^{\dagger}\bw_{k}|^{2}}{\sum_{\breve{k} = 1 , \breve{k} \neq k}^{K} |\bh_{k}^{\dagger}\bw_{\breve{k}}|^{2} + \sigma_{\textrm{C},k}^{2}},
\end{align}
where $\sigma_{\textrm{C},k}^{2}$ is the noise power level in the $k$th communication receiver. For notational simplicity, we assume that $\sigma_{\textrm{C},1}^{2}=\sigma_{\textrm{C},2}^{2}=\cdots=\sigma_{\textrm{C},K}^{2} = \sigma_{\textrm{C}}^{2}$.

\subsection{Radar Model}
Assume that $P$ targets are present, with directions $\theta_1 , \theta_2 , \cdots , \theta_P$.
Under the far-field and narrowband assumption, the signals received by the DFRC system, denoted $\bY_{\textrm{R}} \in \complexC^{N_{\textrm{R}} \times L}$, can be written as
\begin{align}\label{eq:recive matrix}
	\bY_{\textrm{R}} = \sum\limits_{p = 1}^{P}\alpha_p\ba_{p, \textrm{R}}\ba_{p, \textrm{T}}^{\top}\bX + \bN_{\textrm{R}},
\end{align}
where $\alpha_p$ denotes the amplitude of the $p$th target, $\ba_{p , \textrm{R}}  \triangleq \ba_{\textrm{R}}(\omega_p)  \in \complexC^{N_{\textrm{R}} \times 1}$ and $\ba_{p , \textrm{T}} \triangleq \ba_{\textrm{T}}(\omega_p)  \in \complexC^{N_{\textrm{T}} \times 1}$ are the receive and transmit array steering vectors at $\omega_p=\pi\sin(\theta_p)$, which is the spatial frequency of the $p$th target (in radians), and $\bN_{\textrm{R}} \in \complexC^{N_{\textrm{R}} \times L}$ is the receiver noise.
Let $\bm{\alpha} = [\alpha_1 ,\alpha_2 ,\cdots ,\alpha_P]^{\top} \in \complexC^{P \times 1}$,
$\bA_{\textrm{R}} = [\ba_{1 ,\textrm{R}} , \ba_{2 ,\textrm{R}} , \cdots , \ba_{P ,\textrm{R}}]\in \complexC^{N_{\textrm{R}} \times P}$,
$\bA_{\textrm{T}} = [\ba_{1 ,\textrm{T}} , \ba_{2 ,\textrm{T}} , \cdots , \ba_{P ,\textrm{T}}]\in \complexC^{N_{\textrm{T}} \times P}$, and $ 	\bB = \textrm{Diag}(\bm{\alpha})\in \complexC^{P \times P}$.
Then, $\bY_{\textrm{R}}$ can be rewritten as
\begin{align}\label{eq:recive matrix new}
	\bY_{\textrm{R}} = \bA_{\textrm{R}} \bB \bA^{\top}_{\textrm{T}}\bX + \bN_{\textrm{R}}.
\end{align}

Note that an important task of the DFRC system is to estimate the target angles $\{\theta_p\}_{p=1}^P$ (or equivalently, $\{\omega_p\}_{p=1}^P$) from $\bY_{\textrm{R}}$.
We will consider the CRB for the estimates of $\{\omega_p\}_{p=1}^P$ as the accuracy metric.
Assume that the columns of $\bN_{\textrm{R}}$ are independent and identically distributed (i.i.d.) random variables, obeying a circularly symmetric complex Gaussian distribution with zero mean and covariance matrix $\sigma^2_{\textrm{R}}\bI_{N_{\textrm{R}}}$.
It follows from \cite{Li2008MIMOCRB} that the Fisher information matrix (FIM) for the estimation of $[\bm{\omega}^\top, \textrm{Re}(\bm{\alpha}^\top), \textrm{Im}(\bm{\alpha}^\top)]^\top$ is given by
\begin{align}\label{eq:FIM}
	\bF = \frac{2}{\sigma^{2}_{\textrm{R}}}
	\left[\begin{matrix}
		\textrm{Re}(\bF_{11}) & \textrm{Re}(\bF_{12}) & -\textrm{Im}(\bF_{12}) \\
		\textrm{Re}^{\top}(\bF_{12}) & \textrm{Re}(\bF_{22}) & -\textrm{Im}(\bF_{22}) \\
		-\textrm{Im}^{\top}(\bF_{12}) & -\textrm{Im}^{\top}(\bF_{22}) & \textrm{Re}(\bF_{22})
	\end{matrix}\right] ,
\end{align}
where $\bm{\omega} = [\omega_1, \omega_2, \cdots, \omega_P]^\top$,
\begin{equation}\label{eq:F11}
	\begin{aligned}
		\bF_{11} =
        & (\dot{\bA}_{\textrm{R}}^{\dagger}\dot{\bA}_{\textrm{R}})\odot(\bB^{*}\bA_{\textrm{T}}^{\dagger}\bR_{\bX}^{*}\bA_{\textrm{T}}\bB) 
		 +(\dot{\bA}_{\textrm{R}}^{\dagger}\bA_{\textrm{R}})\odot(\bB^{*}\bA_{\textrm{T}}^{\dagger}\bR_{\bX}^{*}\dot{\bA}_{\textrm{T}}\bB) \\
		& +(\bA_{\textrm{R}}^{\dagger}\dot{\bA}_{\textrm{R}})\odot(\bB^{*}\dot{\bA}_{\textrm{T}}^{\dagger}\bR_{\bX}^{*}\bA_{\textrm{T}}\bB) 
		 +(\bA_{\textrm{R}}^{\dagger}\bA_{\textrm{R}})\odot(\bB^{*}\dot{\bA}_{\textrm{T}}^{\dagger}\bR_{\bX}^{*}\dot{\bA}_{\textrm{T}}\bB),
	\end{aligned}
\end{equation}
\begin{equation}\label{eq:F12}
	\begin{aligned}
		\bF_{12} =& (\dot{\bA}_{\textrm{R}}^{\dagger}\bA_{\textrm{R}})\odot(\bB^{*}\bA_{\textrm{T}}^{\dagger}\bR_{\bX}^{*}\bA_{\textrm{T}}) 
		 + (\bA_{\textrm{R}}^{\dagger}\bA_{\textrm{R}})\odot(\bB^{*}\dot{\bA}_{\textrm{T}}^{\dagger}\bR_{\bX}^{*}\bA_{\textrm{T}}) ,
	\end{aligned}
\end{equation}
\begin{align}\label{eq:F22}
	\bF_{22} =(\bA_{\textrm{R}}^{\dagger}\bA_{\textrm{R}})\odot(\bA_{\textrm{T}}^{\dagger}\bR_{\bX}^{*}\bA_{\textrm{T}}) ,
\end{align}
\begin{equation}\label{eq:dA_R}
	\begin{aligned}
		\dot{\bA}_{\textrm{R}} =& [\dot{\ba}_{1 ,\textrm{R}} , \dot{\ba}_{2 ,\textrm{R}} , \cdots , \dot{\ba}_{P ,\textrm{R}}] \\
		=& \left[\frac{\partial \ba_{\textrm{R}}(\omega_1)}{\partial \omega_1} , \frac{\partial \ba_{\textrm{R}}(\omega_2)}{\partial \omega_2} , \cdots  , \frac{\partial \ba_{\textrm{R}}(\omega_P)}{\partial \omega_P}\right] ,
	\end{aligned}
\end{equation}
\begin{equation}\label{eq:dA_T}
	\begin{aligned}
		\dot{\bA}_{\textrm{T}} =& [\dot{\ba}_{1 ,\textrm{T}} , \dot{\ba}_{2 ,\textrm{T}} , \cdots , \dot{\ba}_{P ,\textrm{T}}] \\
		=& \left[\frac{\partial \ba_{\textrm{T}}(\omega_1)}{\partial \omega_1} , \frac{\partial \ba_{\textrm{T}}(\omega_2)}{\partial \omega_2} , \cdots  , \frac{\partial \ba_{\textrm{T}}(\omega_P)}{\partial \omega_P}\right] ,
	\end{aligned}
\end{equation}
and $\bR_{\bX} = \bX\bX^{\dagger} \approx L \bW \bW^{\dagger}$.

The corresponding CRB matrix is
\begin{align}\label{eq:CRB}
	\bC = \bF^{-1}.
\end{align}

Note that $\bC$ has a complicated form. Motivated by \cite{STOICA1989MUSIC}, to simplify the following analysis, we derive the asymptotic CRB as well as an upper bound on it under the condition that $N_{\textrm{R}} \gg 1$.
\begin{Prop} \label{Prop:1}
  Let $\bC_{\bm{\omega}}$ denote the CRB matrix associated with ${\bm{\omega}}$, i.e.,
  \begin{align}\label{eq:CRB_theta}
	\bC_{\bm{\omega}} = \frac{\sigma^{2}_{\textrm{R}}}{2} \bF_{\bm{\omega}}^{-1} ,
\end{align}
where, using the block matrix inversion lemma,
\begin{align}\label{eq:G}
	\bF_{\bm{\omega}} =\textrm{Re}(\bF_{11}) - \widetilde{\bF}_{12}%
	 \widetilde{\bF}_{22}^{-1}
\widetilde{\bF}_{12} ^\top,
\end{align}
$ \widetilde{\bF}_{12} = [\textrm{Re}(\bF_{12}), -\textrm{Im}(\bF_{12})]$, and
\begin{align*}
   \widetilde{\bF}_{22}=
   \begin{bmatrix}
     \textrm{Re}(\bF_{22}) & -\textrm{Im}(\bF_{22}) \\
		-\textrm{Im}^{\top}(\bF_{22}) & \textrm{Re}(\bF_{22})
   \end{bmatrix}.
\end{align*}
{As $N_{\textrm{R}}$ increases, the $p$th diagonal element of $\bC_{\bm{\omega}}$, denoted $\bC_{\bm{\omega}}(p ,p)$,
  approaches}
  \begin{align}\label{eq:C_theta}
	\bC_{\bm{\omega}}(p ,p) =\frac{\sigma^{2}_{\textrm{R}}}{2|\alpha_{p}|^{2}} \left[\frac{1}{12}N_{\textrm{R}}^{3}b_p+N_{\textrm{R}} (\ddot{b}_p -{|\dot{b}_p|^2}{b_p^{-1}})\right]^{-1},
\end{align}
where $b_p = \ba_{p,\textrm{T}}^{\dagger}\bR_{\bX}^{*}\ba_{p,\textrm{T}}$, $\dot{b}_p = {\ba}_{p,\textrm{T}}^{\dagger} \bR_{\bX}^{*} \dot{\ba}_{p,\textrm{T}}$, and $\ddot{b}_p = \dot{\ba}_{p,\textrm{T}}^{\dagger} \bR_{\bX}^{*} \dot{\ba}_{p,\textrm{T}}$.
Moreover, \eqref{eq:C_theta} is upper bounded by
\begin{align}\label{eq:C_theta_bound}
	\bC_{\bm{\omega}}(p ,p) \le  \frac{6\sigma^{2}_{\textrm{R}}}{|\alpha_{p}|^2N_{\textrm{R}}^{3}b_p}.
\end{align}
\end{Prop}
\begin{IEEEproof}
  See Appendix \ref{Apd:A}.
\end{IEEEproof}

\emph{Remark 1:} The results in Proposition \ref{Prop:1} demonstrate that as the number of receive antennas $N_{\textrm{R}}$ grows, the resolvability of the receive antenna array improves, and the cross correlations between the receive array steering vectors at different directions tend to approach zero. As a result, the asymptotic CRB for the angle estimate of each target is irrelevant to the angles of the other targets. This is in contrast to the case of finite $N_{\textrm{R}}$, where the CRB for estimating the angle of one target is dependent on the angles of the other targets.

Observe that  the asymptotic CRB associated with the $p$th target is inversely proportional to the target SNR (i.e., $|\alpha_{p}|^{2}/\sigma^{2}_{\textrm{R}}$). Also, increasing the number of receive antennas reduces the CRB. More important, the transmit beampattern (i.e., $\ba_{\textrm{T}}(\theta)^\dagger \bR_{\bX}^{*} \ba_{\textrm{T}}(\theta)$) plays a crucial rule in determining the angle estimation accuracy of the MIMO system. This underscores the significance of designing the transmit beampattern of the system. Note that existing studies mainly focus on the  design of transmit beampattern to approximate a desired one (see, e.g., \cite{Stoica2007Probing,LiuFan2018DFRC,Liu2020Joint,Tang2020DFRC}). However, the problem of choosing a desired beampattern is rarely addressed. Note from  \eqref{eq:C_theta_bound} that if the SNR of the target at $\theta_p$ is low, then the response of the transmit beampattern at this direction should be high. Otherwise, the angle estimation error will be large. 

\emph{Remark 2:} If only one communication user is present (i.e., $K=1$), then $\bR_{\bX} = L \bw \bw^\dagger$ (thus the transmit waveforms are coherent in this case). As a result, $b_p = L|\ba_{p,\textrm{T}}^\dagger \bw|^2$, $|\dot{b}_p|^2 = L^2 |\ba_{p,\textrm{T}}^\dagger \bw|^2|\dot{\ba}_{p,\textrm{T}}^\dagger \bw|^2$, and $\ddot{b}_p = L |\dot{\ba}_{p,\textrm{T}}^\dagger \bw|^2$. Thus, $\ddot{b}_p -{|\dot{b}_p|^2}{b_p^{-1}} = 0$, $p=1,2, \cdots, P$, which means that the upper bound in  \eqref{eq:C_theta_bound} is tight in this case. \footnote{For the case of $K=1$ and $P=1$ (i.e., one target and one communication user), the CRB for target angle estimation is derived in \cite[Section III.B]{LIU2022CRB}. For sufficiently large $N_{\rm{R}}$, by using the approximation presented in Appendix \ref{Apd:A}, we can observe that the asymptotic CRB for \cite{LIU2022CRB} is identical to the result in Proposition 1.} 

\emph{Remark 3:} Note that the bound in \eqref{eq:C_theta_bound} has a close connection with the result in \cite{STOICA1989MUSIC}, which is established for a conventional array processing model. Let us write the signal model in \eqref{eq:recive matrix new} as
\begin{align}\label{eq:recive matrix3}
	\bY_{\textrm{R}} = \bA_{\textrm{R}} \bZ + \bN_{\textrm{R}},
\end{align}
where $\bZ = \bB \bA^{\top}_{\textrm{T}}\bX$. The model in \eqref{eq:recive matrix3} is standard in conventional array signal processing, but it ignores the angle information in the transmit array of the MIMO system. Based on the model in \eqref{eq:recive matrix3}, it follows from Theorem 4.3 in \cite{STOICA1989MUSIC} that, for sufficiently large $N_\textrm{R}$ and $L$, the CRB for estimating $\omega_p$ is given by
\begin{equation}
  \widetilde{\bC}_{\bm{\omega}}(p ,p) = \frac{6\sigma^{2}_{\textrm{R}}}{N_{\textrm{R}}^{3} \bQ(p,p)},
\end{equation}
where $\bQ(p,p)$ is the $p$th diagonal element of $\bQ$, and $\bQ \approx \bZ\bZ^\dagger$. Since $ \bZ\bZ^\dagger= \bB \bA^{\top}_{\textrm{T}}\bR_{\bX}\bA^{*}_{\textrm{T}}\bB^\dagger$, we have $\bQ(p,p) = |\alpha_{p}|^2b_p$. As a result,
\begin{equation}
  \widetilde{\bC}_{\bm{\omega}}(p ,p) = \frac{6\sigma^{2}_{\textrm{R}}}{N_{\textrm{R}}^{3} |\alpha_{p}|^2b_p},
\end{equation}
which coincides with the upper bound in \eqref{eq:C_theta_bound}. Therefore, if the target angle information in the transmit array of the MIMO system is ignored, the angle estimation errors increase. Nevertheless, doing so also results in a much simpler metric for the target angle estimation performance. 
\subsection{Problem Formulation}
In this section, we aim to design a dual-function beamforming matrix that improves the target angle estimation performance as well as guarantees the QOS for communications. To this end, we  assume that prior knowledge about the target (i.e., the target amplitudes $\{\alpha_p\}_{p=1}^{P}$ and directions $\{\theta_p\}_{p=1}^{P}$) as well as the communication channel response is available. Such an assumption is justified by the fact that prior knowledge about the target can be obtained from previous scans (e.g., the authors in \cite{XuLuzhou2008TAES} proposed several algorithms for MIMO systems to accurately estimate the target directions and amplitudes without any secondary data). Moreover, the communication channel response can be estimated by sending pilot signals. Under the above assumption, we formulate a constrained optimization problem for minimizing the CRB. However, since the expression of the asymptotic CRB in \eqref{eq:C_theta} is still rather complicated and the resultant design problem will have a complex optimization landscape, we employ the upper bound in \eqref{eq:C_theta_bound} instead. Moreover, we enforce an SINR constraint on the signals received by the communication receivers to guarantee the QOS. Specifically, the design problem is stated as follows:
\begin{subequations} \label{eq:P1}
	\begin{align}
		\min_{\bW} &\ \sum_{p=1}^{P}\frac{1}{|\alpha_p|^2\ba_{p{, }\textrm{T}}^{\top}\bW\bW^{\dagger}\ba_{p{, }\textrm{T}}^{*}} \label{eq:P1 pro} \\
		\textrm{s.t.} &\ {\rho_{k}} \geq \hat{\Gamma}_{k} ,  k = 1,  2,  \cdots , K, \label{eq:P1 con1}  \\
		&\ \textrm{tr} (\bW\bW^{\dagger}) \leq e_{\textrm{T}},  \label{eq:P1 con2}
	\end{align}
\end{subequations}
where the objective is proportional to the sum of the CRB upper bounds for the $P$ targets (it is also tantamount to {maximizing} the harmonic mean of the weighted beampattern responses at the target directions),  $\hat{\Gamma}_{k}$ is the minimum SINR that guarantees the communication QOS, $e_{\textrm{T}} = \hat{e}_{\textrm{T}} /L$, and $\hat{e}_{\textrm{T}}$ is the maximum transmit energy.

Note that the SINR constraint in \eqref{eq:P1 con1} is equivalent to
\begin{align}\label{eq:simplify P1 con1}
\textrm{tr}(\bm{\Lambda}_{k} \bW^{\dagger} \hat{\bH}_{k} \bW) \geq \Gamma_{k},
\end{align}
where $\bm{\Lambda}_{k} = \textrm{Diag} ([\underbrace{-\hat{\Gamma}_{k} , \cdots ,-\hat{\Gamma}_{k}}_{k-1} , 1 , \underbrace{-\hat{\Gamma}_{k} , \cdots ,-\hat{\Gamma}_{k}}_{K-k} ]) = (\hat{\Gamma}_{k}+1) \be_{k} \be_{k}^{\top} - \hat{\Gamma}_{k} \bI_{K}$, $\be_{k}$ is the $k$th column of $\bI_{K}$, $\hat{\bH}_{k} = \bh_{k} \bh_{k}^{\dagger}$, and $\Gamma_{k} = \hat{\Gamma}_{k} \sigma_{\textrm{C}}^{2}$. Consequently, we can recast the optimization problem in \eqref{eq:simplify P1 con1} as
\begin{subequations} \label{eq:P2}
	\begin{align}
		\min_{\bW} &\ \sum_{p=1}^{P}\frac{1}{|\alpha_p|^2 \textrm{tr} (\bW^{\dagger}\hat{\bA}_{p}\bW)} \label{eq:P2 pro} \\
		\textrm{s.t.} &\  \textrm{tr}(\bm{\Lambda}_{k} \bW^{\dagger} \hat{\bH}_{k} \bW) \geq \Gamma_{k}, k = 1,  2, \cdots , K, \label{eq:P2 con1}  \\
		&\ \textrm{tr} (\bW\bW^{\dagger}) \leq e_{\textrm{T}}{, } \label{eq:P2 con2}
	\end{align}
\end{subequations}
where $\hat{\bA}_{p} = \ba_{p,\textrm{T}}^{*} \ba_{p,\textrm{T}}^{\top}$.

Using the fact that $\textrm{tr}(\bA \bB \bC \bD) = \textrm{vec}^{\top}(\bD)(
\bA \otimes \bC^{\top} ) \textrm{vec}(\bB^{\top})$ \cite{HornJohnson1990matrixbook}, we have
\begin{align}\label{eq:identity}
	\textrm{tr}(\bm{\Lambda}_{k} \bW^{\dagger} \hat{\bH}_{k} \bW) = \bw^{\dagger} \hat{\bT}_{k} \bw ,  \\
	\textrm{tr} (\bW^{\dagger}\hat{\bA}_{p}\bW) = \bw^{\dagger} \bA_{p} \bw ,
\end{align}
where $\bw = \textrm{vec}({\bW^{*}})$, $\hat{\bT}_{k} = \bm{\Lambda}_{k} \otimes \hat{\bH}_{k}^{\top}$, and $\bA_{p} = \bI_{K} \otimes \hat{\bA}^{\top}_{p}$. Thus, the optimization problem in \eqref{eq:simplify P1 con1} can be rewritten as
\begin{subequations} \label{eq:P3}
	\begin{align}
		\min_{\bw} &\ \sum_{p=1}^{P}\frac{1}{|\alpha_p|^2 \bw^{\dagger} \bA_{p} \bw} \label{eq:P3 pro}\\
		\textrm{s.t.} &\ \bw^{\dagger} \hat{\bT}_{k} \bw \geq \Gamma_{k}, k = 1, 2,  \cdots , K, \label{eq:P3 con1}  \\
		&\ \bw^{\dagger}\bw \leq e_{\textrm{T}}. \label{eq:P3 con2}
	\end{align}
\end{subequations}

Note that the optimal solution to \eqref{eq:P3} must satisfy $\bw^{\dagger}\bw = e_{\textrm{T}}$. Based on this observation, let us define
\begin{align}\label{eq:bT}
	\bT_{k} = \hat{\bT}_k - \beta \bI_{N_{\textrm{T}}K},
\end{align}
where $\beta < \lambda_{\min}(\hat{\bT}_k)$.\footnote{It can be verified that $\lambda_{\min}(\hat{\bT}_k) = \lambda_{\min}(\bm{\Lambda}_{k}) \lambda_{\max}(\hat{\bH}_{k}^{\top}) =-\hat{\Gamma}_{k}\|\bh_{k} \|_2^2$. } It is easy to check that $\bT_{k}\succeq \bm{0}$. In addition, $ \bw^{\dagger} {\bT}_{k} \bw= \bw^{\dagger} \hat{\bT}_{k} \bw - \beta e_{\textrm{T}}$. Thus, the optimization problem in \eqref{eq:P3} can be reformulated as
\begin{subequations} \label{eq:P4}
	\begin{align}
		\min_{\bw} &\ \sum_{p=1}^{P}\frac{1}{|\alpha_p|^2 \bw^{\dagger} \bA_{p} \bw} \label{eq:P4 pro} \\
		\textrm{s.t.} &\ \bw^{\dagger} \bT_{k} \bw \geq \eta_k ,  k = 1,  2, \cdots ,  K, \label{eq:P4 con1}  \\
		&\ \bw^{\dagger}\bw = e_{\textrm{T}}, \label{eq:P4 con2}
	\end{align}
\end{subequations}
where $\eta_{k} = \Gamma_{k} - \beta e_{\textrm{T}}$.

\section{Beamforming Optimization Algorithms}\label{Sec:Design}
Note that the optimization problem in \eqref{eq:P4} is nonconvex due to both the objective and the constraints. To tackle this nonconvex problem, we propose two iterative algorithms: The first one is based on ADMM and the second is based on majorization minimization (we refer to \cite{Boyd2011ADMM} as a tutorial introduction to ADMM, and \cite{Stoica2004Cyclic,Sun2017MM} as introductions to majorization minimization). Next, we derive these  two algorithms in details.

\subsection{ADMM}
By using the variable splitting trick and introducing auxiliary variables $\{\bz_p\}_{p=1}^P$ and $\{\bu_k\}_{k=1}^K$, we  recast the  optimization problem in \eqref{eq:P4} as
\begin{subequations} \label{eq:P7}
	\begin{align}
		\min_{\bw,\{\bz_p\},\{\bu_k\}} &\ \sum\limits_{p = 1}^{P} \frac{1}{|\alpha_p|^2\bz_p^{\dagger}\bz_p} \label{eq:P7 pro} \\
		\textrm{s.t.} &\ \bz_p = \bA^{1/2}_{p} \bw, p = 1, 2, \cdots, P, \label{eq:P7 con1} \\
		&\ \bu_k = \bT_{k}^{1/2} \bw, \bu_{k}^{\dagger} \bu_{k} \geq \eta_k,  k = 1,  2,  \cdots , K, \label{eq:P7 con3}  \\
		&\ \bw^{\dagger}\bw = e_{\textrm{T}}. \label{eq:P7 con4}
	\end{align}
\end{subequations}
The augmented Lagrange function for \eqref{eq:P7} can be written as
\begin{align}\label{eq:The augmented Lagrange function}
	& \L_{\mu}(\bw ,\{\bz_{p}\}, \{\u_p\} , \{\bu_{k}\} , \{\v_{k}\}) \nonumber \\
	= &\ \sum\limits_{p = 1}^{P}  \left[|\alpha_{p}|^{-2}(\bz_p^{\dagger}\bz_{p})^{-1} +\frac{\mu}{2}(\|\bz_p-\bA^{1/2}_{p} \bw +\u_p\|_2^2-\|\u_p\|_2^2)\right] \nonumber \\
	&\ + \frac{\mu}{2} \sum\limits_{k = 1}^{K}(\|\bu_{k}- \bT_{k}^{1/2} \bw +\v_{k}\|_2^2-\|\v_{k}\|_2^2) ,
\end{align}
where $\mu$ is a penalty parameter, and $\{\u_p\}_{p=1}^P$ as well as  $\{\v_k\}_{k=1}^K$ are the Lagrange multipliers associated with the equality constraints in \eqref{eq:P7 con1} and \eqref{eq:P7 con3}, respectively. At the ($r+1$)th iteration of the proposed ADMM algorithm, the following steps are performed sequentially:
\begin{subequations}
	\begin{align}
		&\ \bw_{{r+1}} = \mathop {\textrm{argmin}}\limits_{\bw} \L_{\mu}(\bw , \{\bz_{p , r}\} , \{\u_{p , r}\} , \{\bu_{k,r}\} , \{\v_{k,r}\}) , \label{eq:ADMM1} \\
		&\ \bz_{p ,r+1} = \mathop {\textrm{argmin}} \limits_{\bz_{p}} \L_{\mu}(\bw_{r+1} ,\{\bz_p\} ,\{\u_{p ,r}\} , \{\bu_{k,r}\} , \{\v_{k,r}\}) , \label{eq:ADMM2} \\
		&\ \bu_{k ,r+1} = \mathop {\textrm{argmin}} \limits_{\bu_{k}} \L_{\mu}(\bw_{r+1} ,\{\bz_{p,r+1}\} ,\{\u_{p ,r}\} , \{\bu_{k}\} , \{\v_{k,r}\}) , \label{eq:ADMM3} \\
		&\ \u_{p , r+1} = \u_{p , r}+\bz_{p , r+1}- \bA^{1/2}_{p} \bw_{r+1} , \label{eq:ADMM4} \\
		&\ \v_{k , r+1} = \v_{k , r}+\bu_{k , r+1} - \bT_{k}^{1/2} \bw_{r+1}. \label{eq:ADMM5}
	\end{align}
\end{subequations}

\noindent $\bullet$ \textbf{The solution to \eqref{eq:ADMM1}:}\\
We can write the optimization problem in \eqref{eq:ADMM1} as:
\begin{align}\label{eq:optimization problem ADMM1}
	\min_{\bw} &\ \sum\limits_{p = 1}^{P} \|\bz_{p,r}-\bA^{1/2}_{p} \bw +\u_{p,r}\|_2^2 + \sum\limits_{k = 1}^{K}\|\bu_{k,r} - \bT_{k}^{1/2} \bw +\v_{k,r}\|_2^2 \nonumber \\
	\textrm{s.t.} &\ \bw^{\dagger}\bw = e_{\textrm{T}}.
\end{align}
Let
\begin{equation}
\bA = \sum_{p = 1}^{P} \bA_{p} + \sum_{k = 1}^{K} \bT_{k},
\end{equation}
 and
 \begin{equation}
 \bg_{r} = \sum_{p = 1}^{P} \bA^{1/2}_{p} (\bz_{p,r} + \u_{p,r}) + \sum_{k = 1}^{K} \bT_{k}^{1/2} (\bu_{k,r} + \v_{k,r}).
 \end{equation}
Then, the problem in \eqref{eq:optimization problem ADMM1} can be reformulated as
\begin{align}\label{eq:optimization problem S 2}
	\min_{\bw} &\ \bw^{\dagger}\bA\bw - 2\textrm{Re}(\bg_{r}^{\dagger}\bw)\nonumber \\
	\textrm{s.t.} &\ \bw^{\dagger}\bw = e_{\textrm{T}}.
\end{align}
The optimization problem in \eqref{eq:optimization problem S 2} can be solved by the Lagrange multiplier method. The Lagrangian associated with  \eqref{eq:optimization problem S 2} is:
\begin{equation}
  F(\bw,\varpi) = \bw^{\dagger}\bA\bw - 2\textrm{Re}(\bg_{r}^{\dagger}\bw) + \varpi(\bw^{\dagger}\bw - e_{\textrm{T}}),
\end{equation}
where $\varpi$ is the Lagrange multiplier associated with the equality constraint in \eqref{eq:optimization problem S 2}. Setting the derivative of $F(\bw,\varpi)$ with respect to $\bw$ to zero, we obtain the optimal solution to \eqref{eq:optimization problem S 2}, which is given by
\begin{equation}
  \bw_{r+1} = (\bA + \varpi_{r}\bI)^{-1}\bg_r,
\end{equation}
where $\varpi_{r}$ can be obtained by solving the following equation:
\begin{equation}
  \bg_r^\dagger(\bA + \varpi_{r}\bI)^{-2}\bg_r = e_{\textrm{T}}
\end{equation}
using, for instance, a bisection or a Newton's method (see, e.g., \cite{Tang2022TSP,Li2003RCB} for details).

\noindent $\bullet$ \textbf{The solution to \eqref{eq:ADMM2}:}\\
The optimization problem in \eqref{eq:ADMM2} can be written as (in a decoupled form):
\begin{align}\label{eq:optimization problem z_p 1}
	\min_{\bz_{p}} \frac{1}{|\alpha_{p}|^2\bz_p^{\dagger}\bz_{p}} + \frac{\mu}{2}\|\bz_{p}-\bA^{1/2}_{p} \bw_{r+1} +\u_{p,r}\|_2^2.
\end{align}
Define
\begin{align}\label{eq:b}
	\bb_{p,r} = \bA^{1/2}_{p} \bw_{r+1} - \u_{p,r}.
\end{align}
Then, the optimization problem in \eqref{eq:optimization problem z_p 1} can be rewritten  as
\begin{align}\label{eq:optimization problem z_p 2}
	\min_{\bz_{p}} \frac{1}{|\alpha_p|^2 \|\bz_p\|_2^2} + \frac{\mu}{2}\left[\|\bz_p\|_2^2-2\textrm{Re}(\bb_{p{, }r}^{\dagger}\bz_{p})\right].
\end{align}
According to the  Cauchy-Schwartz inequality, we have that
\begin{equation}
  \textrm{Re}(\bb_{p{, }r}^{\dagger}\bz_{p}) \leq \|\bb_{p{, }r}\|_2 \|\bz_{p}\|_2,
\end{equation}
where the upper bound is achieved if
\begin{align}\label{eq:solution 1}
	\bz_{p} = \chi_{p} \bb_{p{, }r}{, }
\end{align}
for any $\chi_{p} >0$. Substituting \eqref{eq:solution 1} into \eqref{eq:optimization problem z_p 2}, one can verify that it is sufficient to solve the following problem to obtain the solution of \eqref{eq:optimization problem z_p 2}:
\begin{align}\label{eq:optimization problem z_p 3}
	\min_{\chi_{p}} &\ f(\chi_{p}) =|\alpha_p|^{-2}\eta_{p,r}^{-1}\chi_{p}^{-2} + \mu\eta_{p,r}(\chi_{p}^2/2-\chi_{p}) \nonumber \\
	\textrm{s.t.} &\ \chi_{p}>0,
\end{align}
where $\eta_{p,r} = \|\bb_{p,r}\|_2^2 > 0$.
It can be checked that the second-order derivative of $f(\chi_{p})$ with respect to $\chi_{p}$ satisfies
\begin{align}\label{eq:second-order derivative}
	\frac{\textrm{d}^2 f(\chi_{p})}{\textrm{d}\chi_{p}^2}= 6|\alpha_p|^{-2}\eta_{p,r}^{-1}\chi_{p}^{-4}+\mu\eta_{p,r}>0.
\end{align}
Thus, $f(\chi_{p})$ is a convex function. As a result, the optimal solution to the problem in \eqref{eq:optimization problem z_p 3} can be obtained by setting the  derivative of $f(\chi_{p})$ with respect to $\chi_{p}$ to zero, i.e., we need to solve the following quartic equation:
\begin{align}\label{eq:quartic equation}
	\mu \chi_{p}^4 -\mu \chi_{p}^3 -2|\alpha_p|^{-2}\eta_{p,r}^{-2}=0.
\end{align}
Note that  \eqref{eq:quartic equation} can be solved by the Ferrari method \cite{Ferrari}. Also note that by Descartes' rule of signs, the above equation has one positive solution. Denote the solution to \eqref{eq:quartic equation} by $\chi_{p,r}$. Then $\bz_{p{, }r+1}$ can be updated by
\begin{align}\label{eq:update z_p}
	\bz_{p{, }r+1} = \chi_{p,r} \bb_{p{, }r}.
\end{align}

\noindent $\bullet$ \textbf{The solution to \eqref{eq:ADMM3}:}\\
The optimization problem in \eqref{eq:ADMM3} can be formulated as:
\begin{subequations}\label{eq:optimization problem ADMM4}
	\begin{align}
		\min_{\bu_{k}} &\ \|\bu_{k}- \bc_{k,r}\|_2^2 \label{eq:ADMM4 pro}\\
		\textrm{s.t.} &\ \|\bu_{k} \|_{2}^{2} \geq \eta_{k}, \label{eq:ADMM4 con1}
	\end{align}
\end{subequations}
where $\bc_{k,r} = \bT_{k}^{1/2} \bw_{r+1} - \v_{k,r}$.

One can verify that the solution to \eqref{eq:optimization problem ADMM4} is given by (note that $\eta_k >0$ so that its square root exists)
\begin{align}\label{eq:solution to ADMM3}
	\bu_{k,r+1} =
	\begin{cases}
		\bc_{k,r}, & \|\bc_{k,r}\|_2^2 \ge \eta_{k}, \\
		\sqrt{\eta_{k}} \cdot \bc_{k,r}/\|\bc_{k,r}\|_2, & \|\bc_{k,r}\|_2^2 < \eta_{k}.
	\end{cases}
\end{align}

\subsection{MM4MM}
Note that a variational form of $1/(|\alpha_p|^2 \bw^{\dagger} \bA_{p} \bw), p = 1,\cdots, P,$ is given by
\begin{align}
	\frac{1}{|\alpha_p|^2 \bw^{\dagger} \bA_{p} \bw} = \max_{\gamma_p \geq 0} -|\alpha_p|^2 \gamma_p\bw^\dagger \bA_p\bw + 2\sqrt{\gamma_p},
\end{align}
where the maximum is achieved if $\gamma_p = 1/(|\alpha_p|^2 \bw^{\dagger} \bA_{p} \bw)^2$.
Therefore, the optimization problem in \eqref{eq:P4} can be reformulated as a constrained minimax problem as follows:
\begin{subequations} \label{eq:P8}
	\begin{align}
		\min_{\bw} \ \max_{\{\gamma_p\} \geq 0} &\ \sum_{p=1}^{P}\left(-|\alpha_p|^2 \gamma_p\bw^\dagger \bA_p\bw + 2\sqrt{\gamma_p}\right)  \\
		\textrm{s.t.} &\ \bw^{\dagger} \bT_{k} \bw \geq \eta_k, k = 1, 2,  \cdots, K,   \label{Constraint:P8b}\\
		&\ \bw^{\dagger}\bw = e_{\textrm{T}}.
	\end{align}
\end{subequations}
Next, we resort to the Lagrangian to deal with the $K$ inequality constraints in \eqref{Constraint:P8b} and reformulate the optimization problem in \eqref{eq:P8} as
\begin{subequations} \label{eq:P9}
	\begin{align}
		\min_{\bw} \ \max_{\substack{\{\gamma_p\}\geq 0 \\ \{\lambda_k\}\geq 0}} &\ \sum_{p=1}^{P} (-|\alpha_p|^2 \gamma_p\bw^\dagger \bA_p\bw + 2\sqrt{\gamma_p}) 
		 + \sum_{k=1}^{K} \lambda_k (\eta_k - \bw^\dagger \bT_k\bw) \\
		\textrm{s.t.} &\ 
		\bw^{\dagger}\bw = e_{\textrm{T}},
	\end{align}
\end{subequations}
where $\lambda_k \geq 0, k=1, 2, \cdots, K$, are the Lagrange multipliers associated with the inequality constraints in \eqref{Constraint:P8b}.

Define
\begin{align} \label{eq:B}
	\bM = \sum_{p=1}^{P} |\alpha_p|^2 \gamma_p\bA_p + \sum_{k=1}^{K}\lambda_k\bT_k.
\end{align}
Then the objective function in \eqref{eq:P9} can be rewritten as
\begin{align}
	g(\bw, \bgamma, \blambda) = -\bw^\dagger\bM\bw + \sum_{p=1}^{P}2\sqrt{\gamma_p} + \sum_{k=1}^{K}\lambda_k\eta_k,
\end{align}
where $\bgamma = [\gamma_1, \gamma_2, \cdots, \gamma_P]^\top$, and $\blambda = [\lambda_1, \lambda_2, \cdots, \lambda_K]^\top$.

One can verify that $\bM\succeq \bzero$. Thus,  $-\bw^\dagger\bM\bw$ is a concave function of $\bw$. According to a property of  concave functions, a majorized function of $-\bw^\dagger\bM\bw$ is given by
\begin{align} \label{eq:MM1}
	-\bw^\dagger\bM\bw \leq -2\textrm{Re}(\bw_r^\dagger\bM\bw) + \bw_r^\dagger\bM\bw_r,
\end{align}
where $\bw_r$ is the solution at the $r$th iteration. Therefore, a majorizing function of $g(\bw,\bgamma,\blambda)$ can be written as
\begin{align} \label{eq:g_s}
	g_s(\bw,\bgamma,\blambda)
=& -2\textrm{Re}(\bw_r^\dagger\bM\bw) + \bw_r^\dagger\bM\bw_r + \sum_{p=1}^{P}2\sqrt{\gamma_p} 
+ \sum_{k=1}^{K}\lambda_k\eta_k.
\end{align}
As a result, the surrogate problem based on \eqref{eq:g_s} at the $(r+1)$th iteration of the majorization minimization algorithm is formulated as
\begin{subequations} \label{eq:P99}
	\begin{align}
		\min_{\bw}\ \max_{\substack{\{\gamma_p\}\geq 0 \\ \{\lambda_k\}\geq 0}} &\ g_s(\bw,\bgamma,\blambda) \\
		\textrm{s.t.} &\ \bw^{\dagger}\bw = e_{\textrm{T}}.
	\end{align}
\end{subequations}
Note that $g_s(\bw,\bgamma,\blambda)$ is linear with respect to $\bw$. Thus, relaxing the equality constraint in \eqref{eq:P99} with the inequality constraint that $\bw^{\dagger}\bw \leq e_{\textrm{T}}$ does not change the optimal solution. With the relaxation, the above optimization problem with respect to $\bw$ is convex. Note also that $g_s(\bw,\bgamma,\blambda)$ is linear in $\blambda$ and concave in $\bgamma$. By using Sion's minimax theorem \cite{Sion1958minimax}, the relaxed problem of \eqref{eq:P99} is equivalent to the following maximin problem
%
\begin{subequations} \label{eq:P10}
	\begin{align}
		\max_{\substack{\{\gamma_p\}\geq 0 \\ \{\lambda_k\}\geq 0}} \ \min_{\bw} &\ -2\textrm{Re}(\bw_r^\dagger\bM\bw) + \bw_r^\dagger\bM\bw_r + \sum_{p=1}^{P}2\sqrt{\gamma_p} + \sum_{k=1}^{K}\lambda_k\eta_k \\
		\textrm{s.t.} &\ \bw^{\dagger}\bw \leq e_{\textrm{T}}.
	\end{align}
\end{subequations}
The inner minimization problem with respect to $\bw$ is  as follows:
\begin{align} \label{eq:update_w}
	\min_{\bw} &\ -\textrm{Re}(\bw_r^\dagger\bM \bw)\nonumber \\
	\textrm{s.t.} &\ \bw^{\dagger}\bw \leq e_{\textrm{T}}.
\end{align}
The closed-form solution to the above problem is given by
\begin{align} \label{eq:w_opt}
	\bw_{r+1} = \frac{\sqrt{e_{\textrm{T}}}\bM \bw_r}{\|\bM\bw_r\|_2}.
\end{align}
Substituting $\bw_{r+1}$ into \eqref{eq:P10} yields
\begin{align} \label{eq:gamma_lambda}
	\max_{\substack{\{\gamma_p\}\geq 0 \\ \{\lambda_k\}\geq 0}} &\ -2\sqrt{e_{\textrm{T}}} \|\bM \bw_r\|_2 + \bw_r^\dagger\bM\bw_r + \sum_{p=1}^{P}2\sqrt{\gamma_p} + \sum_{k=1}^{K}\lambda_k\eta_k,
\end{align}
which is a convex problem and can be solved using any convex solver such as SDPT3 \cite{toh1999sdpt3}.

\emph{Remark 4:} Though the MM$4$MM algorithm can iterate from an infeasible point and find a feasible solution at convergence (as shown by the numerical results), we discuss a simple procedure to find a feasible solution to \eqref{eq:P8} (which is also feasible for the constraints in \eqref{eq:P3}).
Note that $\lambda_{\max}(\hat{\bT}_k) = \lambda_{\max}(\bm{\Lambda}_{k}) \lambda_{\max}(\hat{\bH}_{k}^{\top}) =\|\bh_{k} \|_2^2$. Thus, $\widetilde{\bT}_k=\hat{\bT}_k - \|\bh_{k} \|_2^2  \bI_{N_{\textrm{T}}K} \preceq \bzero$. Let us consider the following optimization problem:
\begin{align} \label{eq:P11}
  \max_{\bw}\ \|\bw\|_2^2,   \textrm{s.t.} \ \bw^{\dagger} \widetilde{\bT}_{k} \bw \geq \widetilde{\eta}_k, k = 1, 2,  \cdots, K,
\end{align}
where $\widetilde{\eta}_k = \Gamma_{k} - \|\bh_{k} \|_2^2e_{\textrm{T}}$.  Let $\widetilde{\bw}$ denote a feasible solution to \eqref{eq:P11}. Next we show that if $\widetilde{e} = \|\widetilde{\bw}\|_2^2 > e_{\textrm{T}}$, then $\bw_f = \sqrt{e_{\textrm{T}}/\widetilde{e}}\cdot\widetilde{\bw}$ is feasible for \eqref{eq:P3} as well as \eqref{eq:P8}. First, it is checked that $\bw_f^\dagger \bw_f = e_{\textrm{T}}$. Since $\bw^{\dagger} \widetilde{\bT}_{k} \bw \leq 0$ for any $\bw$, we must have $\widetilde{\eta}_k \leq 0$. Thus, $\bw_f^{\dagger} \widetilde{\bT}_{k} \bw_f= e_{\textrm{T}}/\widetilde{e}\cdot\widetilde{\bw}^{\dagger} \widetilde{\bT}_{k} \widetilde{\bw} \geq\widetilde{\eta}_k e_{\textrm{T}}/\widetilde{e} \geq \widetilde{\eta}_k$. Note that $\bw_f^{\dagger} \widetilde{\bT}_{k} \bw_f = \bw_f^{\dagger} \hat{\bT}_{k} \bw_f - \|\bh_{k} \|_2^2e_{\textrm{T}}$. As a result, $\bw_f^{\dagger} \hat{\bT}_{k} \bw_f \geq \Gamma_{k}$, i.e., $\bw_f$ is feasible for \eqref{eq:P3} and \eqref{eq:P8}.

We can use the minorizarion maximization technique to tackle the nonconvex  maximization problem in \eqref{eq:P11}. It is easy to verify that
\begin{equation} \label{eq:surrogate}
  \|\bw\|_2^2 \geq -\|\bw_t\|_2^2 + 2 \textrm{Re}(\bw_t^\dagger \bw),
\end{equation}
where $\bw_t$ is the solution to \eqref{eq:P11} at the $t$th iteration. Thus, the surrogate problem based on \eqref{eq:surrogate} at the $(t+1)$th iteration of  the minorizarion maximization algorithm is formulated as
\begin{align} \label{eq:P12}
  \max_{\bw}&\ \textrm{Re}(\bw_t^\dagger \bw) \nonumber \\
  \textrm{s.t.} &\ \bw^{\dagger} \widetilde{\bT}_{k} \bw \geq \widetilde{\eta}_k, k = 1, 2,  \cdots, K,
\end{align}
which is convex and can be solved by a convex solver. Owing to the ascent property of minorizarion maximization based algorithms (i.e., $\|\bw_{t+1}\|_2^2 \geq \|\bw_t\|_2^2$ in this case), we terminate the algorithm whenever $\|\bw_t\|_2^2 \geq e_{\textrm{T}}$. Lastly, we point out that the proposed minorizarion maximization algorithm can be initialized by an arbitrarily chosen vector. Denote the initial point by $\bw_0$. Let $\zeta_k = \bw_0^{\dagger} \widetilde{\bT}_{k} \bw_0, k = 1, 2,  \cdots, K$. If $\zeta_k/\widetilde{\eta}_k \leq 1, k = 1, 2,  \cdots, K$, then $\bw_0$ is feasible. Otherwise, denote $m_k = \max_{k} \zeta_k/\widetilde{\eta}_k$. It is easy to verify that $\bw_0/\sqrt{m_k}$ satisfies all the constraints in \eqref{eq:P12}.
\hfill $\blacksquare$\par

\subsection{Algorithm summary}

\begin{algorithm}[!htbp]
	\DontPrintSemicolon
	\caption{ \small Algorithms for designing dual-function beamforming matrix.} \label{alg1}
	\KwIn{$\{\alpha_p,\ba_{p,\textrm{T}}\}_{p=1}^P, \bH ,e_{\textrm{T}}, \{ \hat{\Gamma}_{k} \}_{k=1}^{K} , \sigma_{\textrm{C}}^{2} ,\mu,\vartheta$.}
	\KwOut{$\bW_{\textrm{opt}}$.}
	\textbf{Initialize:} $r = 0, \bw_{r},  \{\u_{p,r} , \bz_{p ,r} \}_{p=1}^P, \{\v_{k,r} , \bu_{k ,r}\}_{k=1}^K$. \\
	\textbf{Compute:} $\Gamma_{k}$, $\hat{\bH}_{k}$, $\bT_{k}$,	$\eta_{k}$,	$\bA_{p}$, $\bA$.\\
	\Repeat{convergence}{
		\uCase{ADMM Algorithm}{
		Compute $\bg_{r}$.\\
		Update $\bw_{r+1}$ by solving the optimization problem in \eqref{eq:optimization problem S 2}. \\
		\For{$p=1$ to $P$}{
			Compute $\bb_{p,r}$ using \eqref{eq:b}.\\
			Compute $\chi_{p,r}$ in \eqref{eq:quartic equation} using Ferrari method. \\
			$\bz_{p,r+1} = \chi_{p,r} \bb_{p,r}$.\\
		}
		\For{$k=1$ to $K$}{
			$\bc_{k,r} = \bT_{k}^{1/2} \bw_{r+1} - \v_{k,r}$.\\
			$\bu_{k,r+1} = \max(\sqrt{\eta_{k}}/\|\bc_{k,r}\|, 1) \cdot  \bc_{k,r}$.\\
		}
		Update $\u_{p , r+1}$ and $\v_{k , r+1}$ by \eqref{eq:ADMM4} and \eqref{eq:ADMM5}. \\
		$r = r+1$.
		}
		\uCase{MM$4$MM Algorithm}{
			Compute $\bM_{r+1}$. \\
			$\bw_{r+1} = \sqrt{e_{\textrm{T}}} \bM_{r+1}\bw_r/\|\bM_{r+1}\bw_r\|_2$. \\
			Update $\bgamma$ and $\blambda$ by solving \eqref{eq:gamma_lambda}.\\
			$r = r + 1$.
			}

	}$\bW_{\textrm{opt}}= \textrm{unvec}_{N_{\textrm{T}} , K}(\bw_{r}^{*})$.
\end{algorithm}
We summarize the proposed ADMM and MM$4$MM algorithms in Algorithm \ref{alg1}. The iterations are terminated if the following stopping criterion is satisfied:
\begin{align}\label{eq:stop criterion}
	\frac{| h(\bw_{r+1})-h(\bw_{r})|}{h(\bw_{r})} \le \vartheta,
\end{align}
where $h(\bw_{r}) = \sum_{p=1}^{P} 1/{(|\alpha_p|^2 \bw^{\dagger}_{r} \bA_{p} \bw_{r})}$ is the objective value at the $r$th iteration, and $\vartheta$ is the stopping threshold (e.g., $10^{-3}$).

Next, we analyze the computational complexity of the two algorithms. For the ADMM algorithm, at each iteration,
the update of $\bw_{r+1}$ requires $\O((P+K)K^2N_{\textrm{T}}^2)$ flops;
the update of $\{\bz_{p, r+1}\}_{p=1}^{P}$  $\O(PK^{2}N_{\textrm{T}}^{2})$ flops;
the update of $\{\u_{p, r+1}\}_{p=1}^{P}$  $\O(PK^{2}N_{\textrm{T}}^{2})$ flops;
the update of $\{\bu_{k, r+1}\}_{k=1}^{K}$  $\O(K^{3}N_{\textrm{T}}^{2})$ flops;
and the update of $\{\v_{k, r+1}\}_{k=1}^{K}$ $\O(K^{3}N_{\textrm{T}}^{2})$ flops.
Therefore, the total computational complexity of the ADMM algorithm is $\O(N_{\textrm{A}}(3P +3K)K^{2}N_{\textrm{T}}^{2}))$, where $N_{\textrm{A}}$ is the number of iterations needed to reach the convergence.
For the MM$4$MM algorithm, at each iteration,
the update of $\bM$ requires $\O((P+K)K^2N_{\textrm{T}}^2)$ flops;
the update of $\bw_{r+1}$ $\O(N_{\textrm{T}}^2 K^2)$ flops;
and the update of $\bgamma_{r+1}$ and $\blambda_{r+1}$ $\O((K + P)^{3.5})$ flops.
Therefore, the total computational complexity of the MM$4$MM algorithm is $\O(N_{\textrm{M}}((P +K)^{3.5} + (P+K)K^{2}N_{\textrm{T}}^{2}))$, where $N_{\textrm{M}}$ is the number of iterations needed to reach the convergence.

Finally, we note that the proposed algorithm can be extended to design a dual-function beamforming matrix under other constraints.
For example, to reduce the hardware complexity and cost, analog beamforming (also called phase-only beamforming) is of particular interest in MIMO systems \cite{Ayach2014Hybrid}. In analog beamforming, the number of power amplifiers is significantly reduced and only phase-shifters are used to control the beam. In such a case, it is essential to enforce a constant-modulus constraint  on the beamforming matrix, which can be written as
\begin{equation} \label{eq:CMC}
  |w_{n}|=a_\textrm{s},
\end{equation}
where $w_{n}$ is the $n$th element of $\bw$, $n=1,2,\cdots, N_\textrm{T}K$, and $a_\textrm{s}=\sqrt{e_\textrm{T}/(N_\textrm{T}K)}$. To extend the proposed ADMM algorithm to deal with this constraint, we only need to replace the optimization problem in \eqref{eq:optimization problem S 2} by the following
\begin{align}\label{eq:optimization problem analog}
	\min_{\bw} &\ \bw^{\dagger}\bA\bw - 2\textrm{Re}(\bg_{r}^{\dagger}\bw)\nonumber \\
	\textrm{s.t.} &\ |w_{n}|=a_\textrm{s},n=1,2,\cdots, N_\textrm{T}K.
\end{align}
The above optimization problem can be tackled by means of a majorization-minimization algorithm (see, e.g., \cite{Tang2021TSP,Tang2021Information} for details). For the  MM$4$MM algorithm, we only need to replace the optimization problem in \eqref{eq:update_w} by the following one:
\begin{align} \label{eq:update_w_CM}
	\min_{\bw} &\ -\textrm{Re}(\bw_r^\dagger\bM \bw)\nonumber \\
	\textrm{s.t.} &\ |w_{n}| \leq a_\textrm{s},n=1,2,\cdots, N_\textrm{T}K,
\end{align}
where we have relaxed the equality constraint in \eqref{eq:CMC} with the convex inequality constraint $|w_{n}| \leq a_\textrm{s}$.
Note that this relaxation does not change the optimal solution, which is given by
\begin{equation}\label{eq:Solution_CMC}
	w_{n} = a_\textrm{s} \textrm{exp}(\textrm{j}\arg((\bM \bw_r)_{n})),
\end{equation}
where $(\bM \bw_r)_{n}$ denotes the $n$th element of $\bM \bw_r$.
With the result in \eqref{eq:Solution_CMC}, the surrogate problem  at the $(r+1)$th iteration of the MM$4$MM algorithm is given by
\begin{align*} 
	\max_{\substack{\{\gamma_p\}\geq 0 \\ \{\lambda_k\}\geq 0}} &\ -2a_\textrm{s} \|\bM \bw_r\|_1 + \bw_r^\dagger\bM\bw_r + \sum_{p=1}^{P}2\sqrt{\gamma_p} + \sum_{k=1}^{K}\lambda_k\eta_k,
\end{align*}
which is convex and thus can be solved via a convex solver.

\section{Numerical Results}\label{Sec:Numerical Results}
In this section, we provide several numerical examples to demonstrate the performance of the proposed algorithms. Unless otherwise stated, we assume that the MIMO DFRC system is equipped with $N_{\textrm{T}} = 16$ transmit antennas and $N_{\textrm{R}} = 20$ receive antennas. Both the antenna arrays are ULAs with inter-element spacing of half wavelength, and $e_{\textrm{T}} = 0$ dB.
The noise powers in  the communication receivers and in the DFRC system are $\sigma_{\textrm{C}}^{2} = \sigma_{\textrm{R}}^{2} = 0$ dBm.
We assume a flat fading communication channel. Specifically, the elements of $\bH$ are i.i.d. and they obey a Gaussian distribution with zero mean and unit variance. The data streams transmitted to the communication users are $16$-quadrature amplitude modulated ($16$QAM) signals with code length $L = 30$.
There are $P=2$ targets at the directions $\theta_{1} = -5^{\circ}$ and $\theta_{2} = 15^{\circ}$. The target amplitudes are $|\alpha_{1}|^{2} =|\alpha_{2}|^{2} = 0$ dB.
In addition, the MIMO DFRC system serves $K = 6$ communication users.
The SINR threshold that guarantees the communication QOS is $\hat{\Gamma}_{k} = 15$ dB ($k = 1 , 2 , \cdots , K$).
In the proposed algorithms, the beamforming vector $\bw_{0}$ as well as the Lagrange multipliers $\{\u_{p}\}_{p=1}^P$ and $\{\v_{k}\}_{k=1}^K$ are randomly initialized. The penalty parameter in the ADMM algorithm is $\mu = 0.86$.
The CVX toolbox \cite{grant2014cvx} is used to solve the optimization problem in \eqref{eq:gamma_lambda}.
The threshold of the stopping criterion is $\vartheta = 10^{-3}$.
Finally, all the experiments are conducted on a standard laptop with Intel(R) Core(TM) i7-9750H CPU and 16 GB memory.

\begin{figure*}[!htbp]
	\setlength{\abovecaptionskip}{0.cm}
	\setlength{\abovecaptionskip}{0.cm}
	\setlength{\belowdisplayskip}{0pt}
	\centering
{\subfigure[]{{\includegraphics[width = 0.42\textwidth]{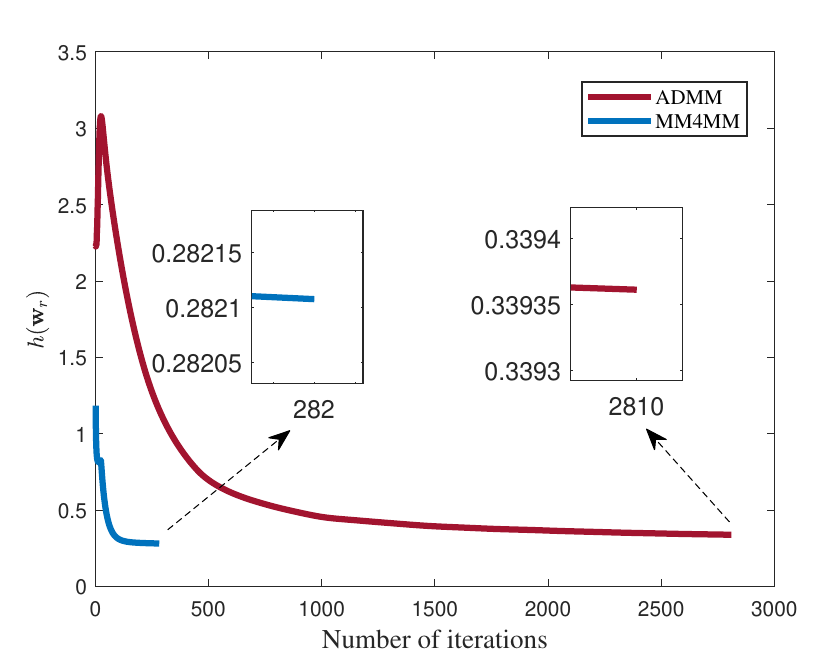}} \label{Fig:2a}} }
{\subfigure[]{{\includegraphics[width = 0.42\textwidth]{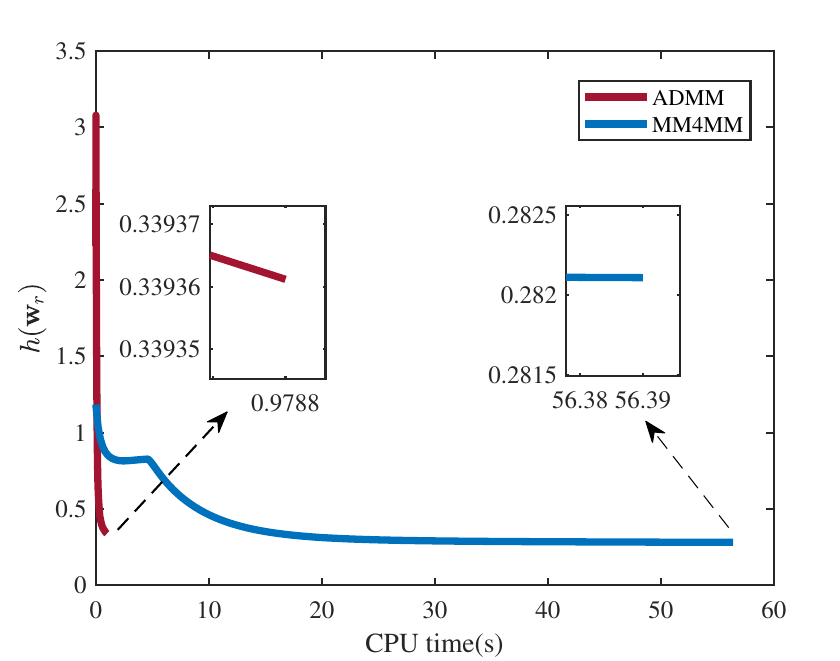}} \label{Fig:2b}} }
	\caption{$h(\bw_{r})$ versus the number of iterations and versus the CPU time. $P = 2$. $|\alpha_{1}|^{2} =|\alpha_{2}|^{2} = 0$ dB. $K =6$. $\hat{\Gamma}_{k} = 15$ dB ($k=1,2,\cdots,K$).}
	\label{Convergence_value}
\end{figure*}

\figurename~\ref{Convergence_value} shows the objective of the optimization problem in \eqref{eq:P4} (i.e., $h(\bw_{r})$) versus the number of iterations and versus the CPU time for the two proposed algorithms.
Observe that the ADMM algorithm requires a larger number of iterations than the MM$4$MM algorithm to converge and has a larger objective value at convergence. However, for the MM$4$MM algorithm, invoking CVX at each iteration (to solve \eqref{eq:gamma_lambda}) is time-consuming, resulting in a longer CPU time to reach convergence than the ADMM algorithm.

\begin{figure}[!htbp]
	\setlength{\abovecaptionskip}{0.cm}
	\setlength{\abovecaptionskip}{0.cm}
	\setlength{\belowdisplayskip}{0pt}
	\centering
	\includegraphics[width= 0.42 \textwidth] {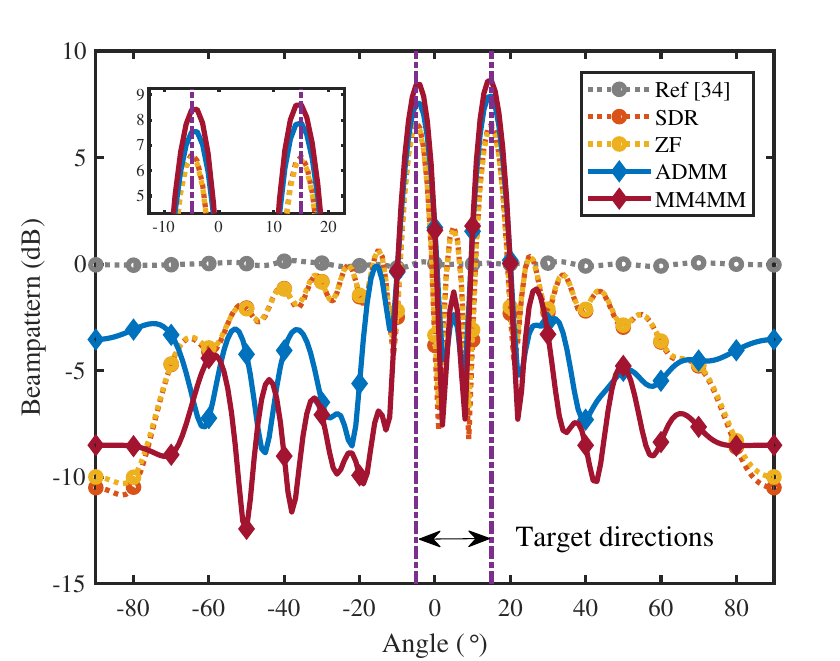}
	\caption{Transmit beampatterns. $P = 2$. $|\alpha_{1}|^{2} =|\alpha_{2}|^{2} = 0$ dB. $K =6$. $\hat{\Gamma}_{k} = 15$ dB ($k=1,2,\cdots,K$).}
	\label{Beampattern}
\end{figure}

\figurename~\ref{Beampattern} shows the transmit beampatterns associated with the beamforming matrices designed by the proposed algorithms and those designed by the algorithms in \cite{Liu2020Joint} and {\cite[Section IV]{LIU2022CRB}}. Note that the algorithm in \cite{Liu2020Joint} aims to minimize a linear combination of the beampattern matching error and the mean-squared cross correlation under per-antenna power constraints as well as a communication SINR constraint for each communication user. To ensure fair comparisons, we minimize the beampattern matching error under the communication SINR constraint and the transmit energy constraint, where the desired beampattern is given by
\begin{equation}
  d(\theta)=
  \begin {cases}
     1, & \theta_p - \frac{\Delta}{2} \leq \theta \leq \theta_p + \frac{\Delta}{2}, p=1,\cdots,P,\\
     0, & \textrm{otherwise},
  \end {cases}
 \end{equation}
and $\Delta=4^\circ$. The transmit beampattern is defined as\footnote{$\ba_{\textrm{T}}(\theta)$ is the transmit array steering vector at $\theta$, defined similarly to $\ba_{\textrm{T}}(\omega)$.}
\begin{align}\label{eq:beampattern}
	P(\theta) = \ba_{\textrm{T}}^{\top}(\theta)\bW \bW^{\dagger} \ba_{\textrm{T}}^{*}(\theta).
\end{align}
Since the prior knowledge about the targets is employed in the proposed algorithms as well as  the semidefinite relaxation (SDR) algorithm and the zero-forcing (ZF) algorithm in \cite{Liu2020Joint}, the beampatterns of the beamforming matrices designed by these algorithms have two mainlobes at the target directions.
Therefore, the transmit energy is focused in the target directions.
The beampattern responses associated with the ADMM algorithm and the MM$4$MM algorithm at the target directions are slightly stronger than that of the SDR algorithm and the ZF algorithm,
{implying that the ADMM algorithm and the MM$4$MM algorithm will achieve lower CRB than the SDR and ZF algorithms.}
In addition, the sidelobes associated with the ADMM algorithm and the MM$4$MM algorithm are lower than those corresponding to the SDR algorithm and the ZF algorithm. In contrast to our designs, the prior knowledge about the targets is not incorporated in the design metric of {\cite[Section II.C]{LIU2022CRB}}, resulting in that the beampattern of the designed beamforming matrix is almost omnidirectional, which will result in energy dispersion.

\begin{figure}[!htbp]
	\setlength{\abovecaptionskip}{0.cm}
	\setlength{\abovecaptionskip}{0.cm}
	\setlength{\belowdisplayskip}{0pt}
	\centering
	\includegraphics[width= 0.42 \textwidth] {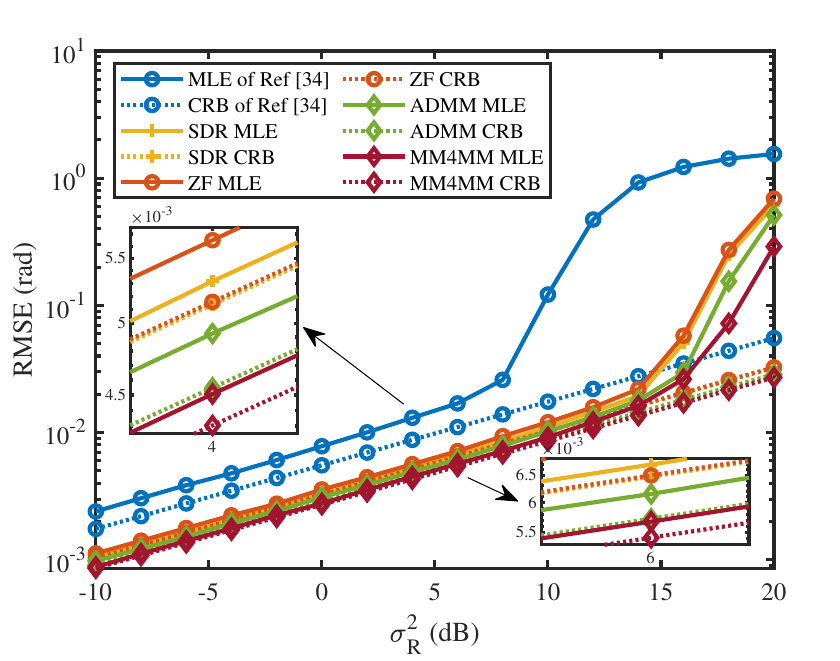}
	\caption{RMSEs versus $\sigma_{\textrm{R}}^{2}$. $P = 2$. $|\alpha_{1}|^{2} =|\alpha_{2}|^{2} = 0$ dB. $K =6$. $\hat{\Gamma}_{k} = 15$ dB ($k=1,2,\cdots,K$).} \label{CRB_RMSE}
\end{figure}

\figurename~\ref{CRB_RMSE} plots the root-CRB (computed using the exact expression in \eqref{eq:CRB}) and the root mean square error (RMSE) versus the noise power for the beamforming matrices corresponding to  \figurename~\ref{Beampattern}. The  RMSE of the spatial frequency estimates is defined as
\begin{align}\label{eq:RMSE}
	\textrm{RMSE}  = \sqrt{\sum_{j=1}^{J} \sum_{p=1}^{P} (\hat{\omega}_{j,p} - \omega_{p})^{2}/J},
\end{align}
where $J = 5000$ is the number of independent Monte Carlo trials, and $\hat {\omega}_{j,p}$ is the maximum likelihood estimate (MLE) of  $\omega_{p}$ in the $j$th trial.  Because the receive array is a ULA, we can use the method of direction estimation (MODE) \cite{Stoica1990MLE,Li1998Comparative} to efficiently obtain the MLE of the target angles.
As shown in  \figurename~\ref{CRB_RMSE}, the proposed MM4MM algorithm reaches the lowest CRB and RMSE (slightly lower than those of the ADMM algorithm).
Moreover, for all the designs that can direct the transmit energy toward the targets, the RMSE curves of the MLE are closer to their corresponding CRB than the design proposed in \cite{LIU2022CRB}. Thus, by incorporating the prior knowledge about the targets into the design of beamforming matrix, the target angle estimates can be refined. {In addition, through minimizing an upper bound on the asymptotic CRB, the beamforming matrices designed by the proposed algorithms achieve lower estimation errors than that  designed based on minimizing the beampattern matching error (i.e., the beamforming matrices  designed by the SDR algorithm and the ZF algorithm). Interestingly, although the feasibility region associated with the ZF algorithm is smaller than that of the SDR algorithm (due to additional constraints), the curves in \figurename~\ref{Beampattern}  and \figurename~\ref{CRB_RMSE} demonstrate that the beampattern matching error as well as the RMSE of the ZF algorithm is only slightly larger than those of the SDR algorithm.}


Table \ref{Table:SINR_threshold} presents the SINR of the communication signals  received by the $K$ users. We can observe that all the designs satisfy the communication SINR constraint, verifying the feasibility of the designed beamforming matrices. For the beamforming matrix designed by \cite{LIU2022CRB},
the SINR of the received communication signals is identical to the threshold that guarantees the communication performance (i.e., $\hat{\Gamma}_{k}$). On the other hand, for the beamforming matrices designed by the proposed algorithms and the SDR algorithm, we can observe  that the SINRs of the received communication signals are slightly higher than the threshold.
{Note that by introducing additional constraints, the inter-user interference associated with the ZF algorithm can be significantly reduced. Thus, the corresponding SINR of the received communication signals is much higher than that of the other algorithms. }

\begin{figure*}[!htbp]
	\setlength{\abovecaptionskip}{0.cm}
	\setlength{\abovecaptionskip}{0.cm}
	\setlength{\belowdisplayskip}{0pt}
	\centering
	\includegraphics[width= 0.8 \textwidth] {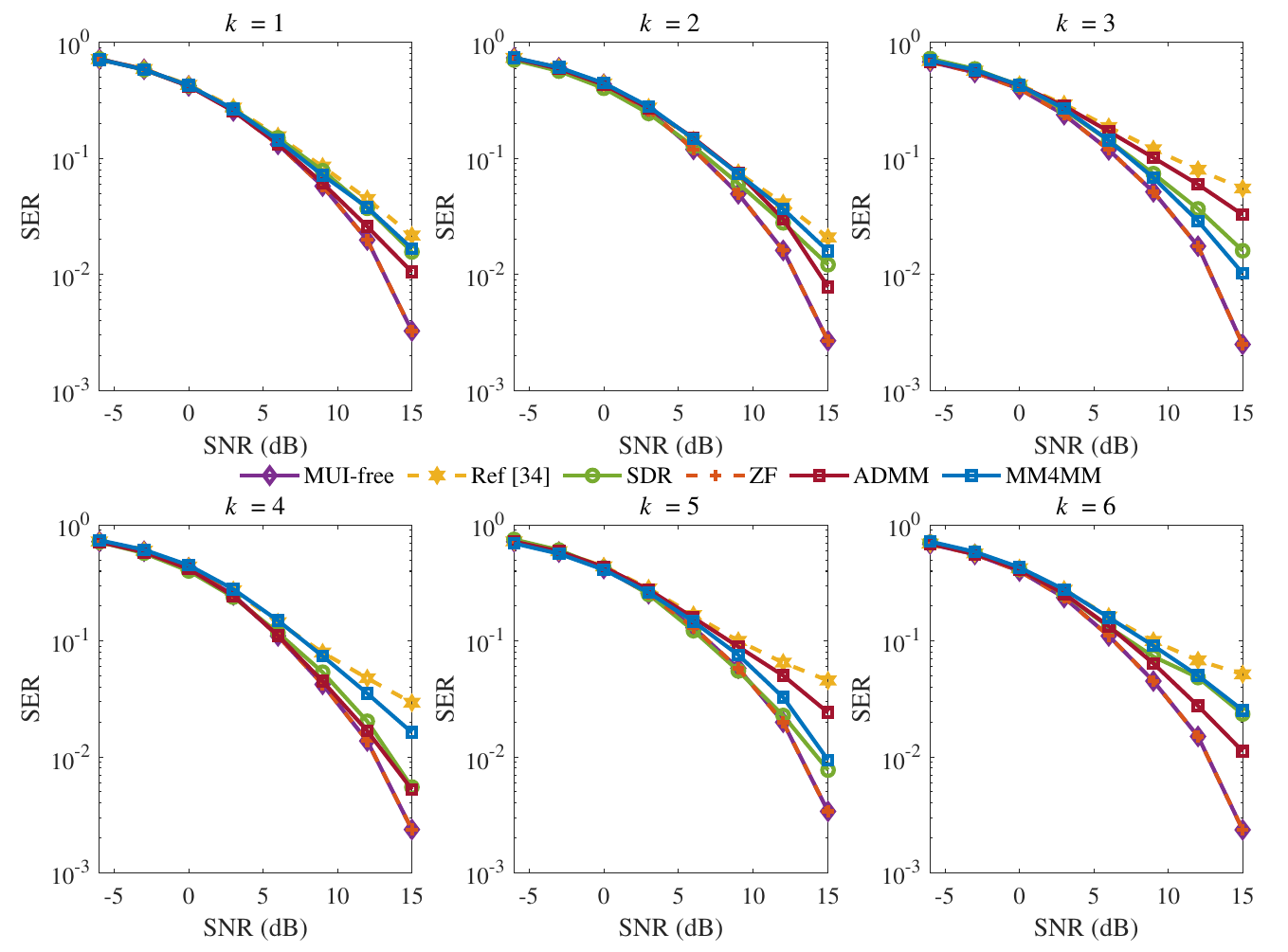}
	\caption{SER for each user. $P = 2$. $|\alpha_{1}|^{2} =|\alpha_{2}|^{2} = 0$ dB. $K =6$. $\hat{\Gamma}_{k} = 15$ dB ($k=1,2,\cdots,K$).}
	\label{SER}
\end{figure*}

\figurename~\ref{SER} analyzes the symbol error rate (SER) of the received communication signals  versus the signal to noise ratio (SNR). The SNR of the $k$th communication signal is defined by
\begin{align}\label{eq:communication SNR}
	\textrm{SNR}_{k}  = \frac{\mathbb{E} \{ |s_{k , l}|^{2}\}}{\sigma_{\textrm{C}}^{2}},
\end{align}
where $s_{k , l}$ is the $l$th symbol of $\bs_{k}$ ($l = 1$, $2$, $\cdots$, $L$). {The performance of the MUI-free case (which corresponds to  a single user system in additive white Gaussian noise with the same SNR) is included as a benchmark}. We conduct $5000$ Monte Carlo trials to obtain the SER. Since the ZF algorithm achieves the highest SINR, the corresponding SER is almost identical to that of {the MUI-free case}. Moreover, owing to the slightly higher SINR, the SERs of the received communication signals for the proposed designs as well as the SDR algorithm are lower than that for the design in \cite{LIU2022CRB}.

\begin{table}[!htbp]
	\caption{{{Communication SINR for each user.}}}
	\centering
	\resizebox{0.5\textwidth}{!}{
  \begin{tabular}{c| c| c| c| c| c| c}
	\bottomrule[1.5pt]
	\thead{Communication \\ SINR (dB)} & $k=1$ & $k=2$ & $k=3$ & $k=4$ & $k=5$ & $k=6$ \\
	\hline
	Ref [34] & 15 & 15 & 15 & 15 & 15 & 15  \\
	\hline
	SDR & 15.049 & 15.093 & 15.030 &15.037  & 15.279 & 15.237  \\
	\hline
	ZF & 25.013 & 24.183 & 23.310 & 20.319 & 29.632 & 24.014 \\
    \hline
	ADMM & 15.230 & 15.232 & 15.017 & 15.112 & 15.046 & 15.318 \\
	\hline
	MM$4$MM  & 15.037 & 15.077 & 15.139 & 15.009 & 15.124 & 15.053  \\
   \toprule[1.5pt]
	\end{tabular}
	}
	\label{Table:SINR_threshold}
  \end{table}

\begin{figure}[!htbp]
	\setlength{\abovecaptionskip}{0.cm}
	\setlength{\abovecaptionskip}{0.cm}
	\setlength{\belowdisplayskip}{0pt}
	\centering
	\includegraphics[width= 0.42 \textwidth] {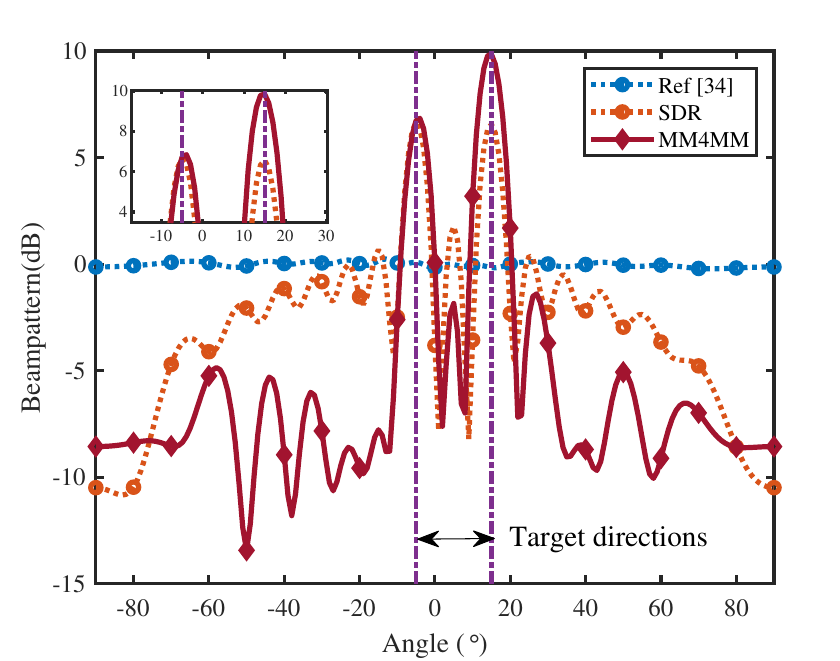}
	\caption{Transmit beampatterns. $P = 2$. $|\alpha_{1}|^{2} =3$ dB and $|\alpha_{2}|^{2} = -3$ dB. $K =6$. $\hat{\Gamma}_{k} = 15$ dB ($k=1,2,\cdots,K$).}
	\label{Beampattern_Unequal}
\end{figure}

\begin{figure}[!htbp]
	\setlength{\abovecaptionskip}{0.cm}
	\setlength{\abovecaptionskip}{0.cm}
	\setlength{\belowdisplayskip}{0pt}
	\centering
	\includegraphics[width= 0.42 \textwidth] {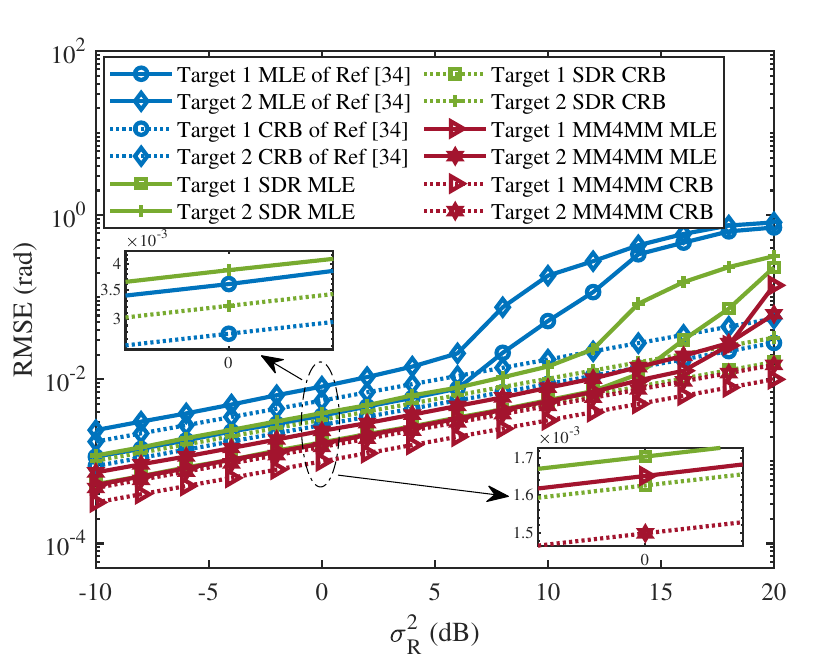}
	\caption{RMSEs versus $\sigma_{\textrm{R}}^{2}$. $P = 2$. $|\alpha_{1}|^{2} =3$ dB and $|\alpha_{2}|^{2} = -3$ dB. $K =6$. $\hat{\Gamma}_{k} = 15$ dB ($k=1,2,\cdots,K$).} \label{CRB_RMSE_Unequal}
\end{figure}

Next we consider a case in which the target amplitudes are not identical.
We use the same parameter setting as in \figurename~\ref{Beampattern}, but now with $|\alpha_{1}|^{2} =3$ dB and $|\alpha_{2}|^{2} = -3$ dB.
Since the performance of the MM$4$MM algorithm is better than that of the ADMM algorithm (but at the cost of longer running time), we only present the result associated with the MM$4$MM algorithm in the sequel to avoid cluttering the figures.
Moreover, as the SDR algorithm achieves slightly better angle estimation performance than the ZF algorithm, we do not include the results associated with the ZF algorithm hereinafter.
\figurename~\ref{Beampattern_Unequal} compares  the transmit beampattern of the beamforming matrix designed by the MM$4$MM algorithm with that by the SDR algorithm and the algorithm in \cite{LIU2022CRB}. Though the parameter setting is different from that in \figurename~\ref{Beampattern}, the beampattern of the beamforming matrix in \cite{LIU2022CRB} is still nearly omnidirectional.
The beampattern of the beamforming matrix designed by the SDR algorithm in \cite{Liu2020Joint} forms two mainlobes at the target directions, but with equal peak response.
{For the proposed design, the beampattern response at target $2$ is stronger than at target $1$ because the SNR of target $2$ is lower.} Moreover, the sidelobes of the proposed design are lower than those of the competing design. \figurename~\ref{CRB_RMSE_Unequal} shows the RMSE of the spatial frequency estimates for each target versus the noise power. The associated CRB curves are also included as a benchmark.
Once again, the designs directing the transmitting energy toward the targets achieve lower CRB and RMSE than the omnidirectional design in \cite{LIU2022CRB}. Since the proposed design achieves a higher  beampattern response for target $2$ (i.e., the weaker target), the corresponding RMSE and the CRB for angle estimate of target $2$ is visibly lower than the other designs.

\begin{figure}[!htbp]
	\setlength{\abovecaptionskip}{0.cm}
	\setlength{\abovecaptionskip}{0.cm}
	\setlength{\belowdisplayskip}{0pt}
	\centering
	\includegraphics[width= 0.42 \textwidth] {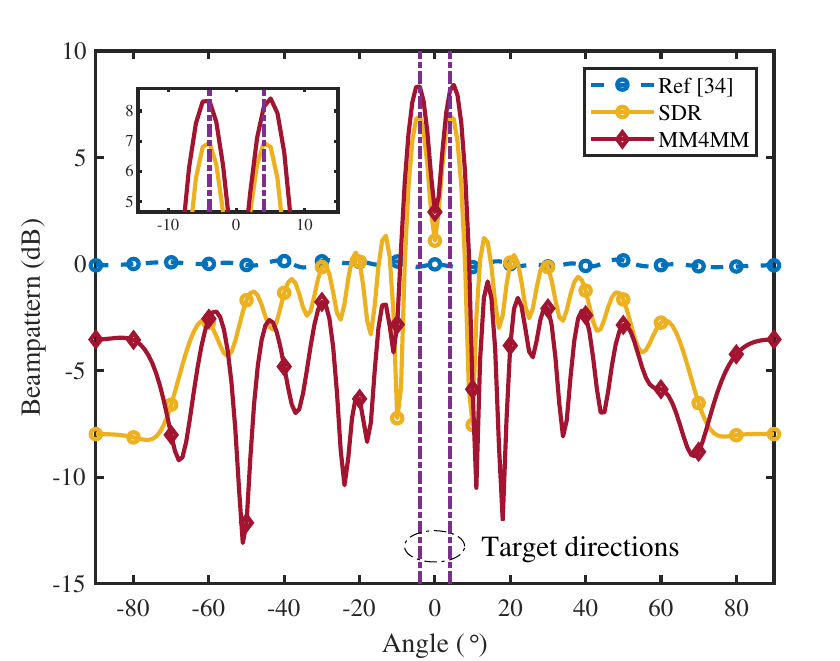}
	\caption{Transmit beampatterns. $P = 2$. $|\alpha_{1}|^{2} = |\alpha_{2}|^{2} = 0$ dB. $\theta_1 = -4^\circ$. $\theta_2 = 4^\circ$. $K =6$. $\hat{\Gamma}_{k} = 15$ dB ($k=1,2,\cdots,K$).}
	\label{Beampattern_8}
\end{figure}

\begin{figure}[!htbp]
	\setlength{\abovecaptionskip}{0.cm}
	\setlength{\abovecaptionskip}{0.cm}
	\setlength{\belowdisplayskip}{0pt}
	\centering
	\includegraphics[width= 0.42 \textwidth] {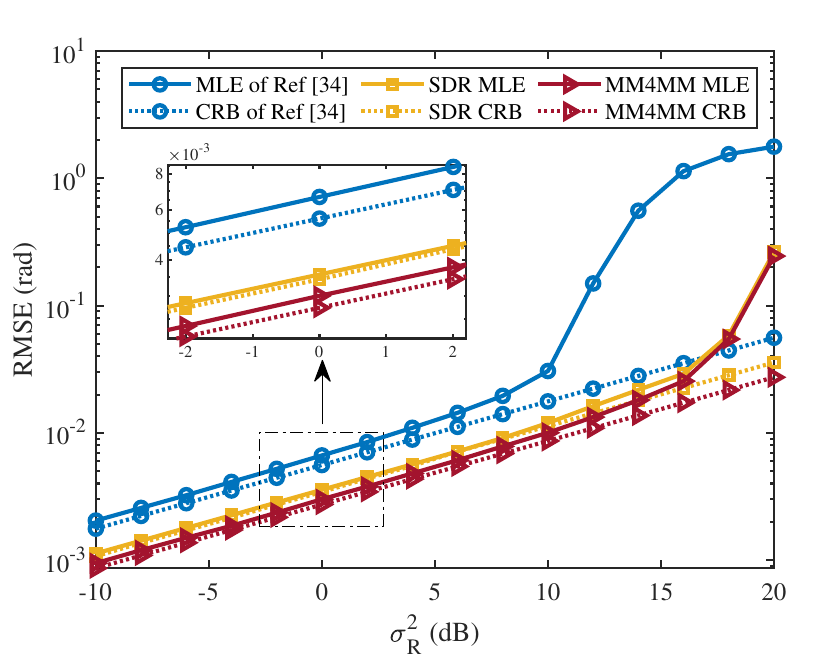}
	\caption{RMSEs versus $\sigma_{\textrm{R}}^{2}$. $P = 2$. $|\alpha_{1}|^{2} = |\alpha_{2}|^{2} = 0$ dB. $\theta_1 = -4^\circ$. $\theta_2 = 4^\circ$. $K =6$. $\hat{\Gamma}_{k} = 15$ dB ($k=1,2,\cdots,K$).} \label{RMSE_AngularSpacing_8}
\end{figure}

{Now} we consider a case in which the two target are closely spaced. We use the same parameter setting as in \figurename ~\ref{Beampattern}, but now with $\theta_1 = -4^\circ$ and $\theta_2 = 4^\circ$. \figurename ~\ref{Beampattern_8} compares the transmit beampattern of the beamforming matrices designed by the three algorithms. Observe that the new parameter setting does not change the  beampattern shape of the beamforming matrix designed by the algorithm in \cite{LIU2022CRB}. For the beamforming matrix designed by the MM$4$MM algorithm and the SDR algorithm, the two mainlobes of the transmit beampattern become closer. Moreover, for the proposed design, the responses at the target directions are higher and the sidelobes are lower. \figurename ~\ref{RMSE_AngularSpacing_8} shows the RMSE and CRB for the spatial frequency estimates of the two targets. It can be seen that the proposed MM$4$MM algorithm reaches the lowest CRB and RMSE. 

\begin{figure}[!htbp]
	\setlength{\abovecaptionskip}{0.cm}
	\setlength{\abovecaptionskip}{0.cm}
	\setlength{\belowdisplayskip}{0pt}
	\centering
	\includegraphics[width= 0.42 \textwidth] {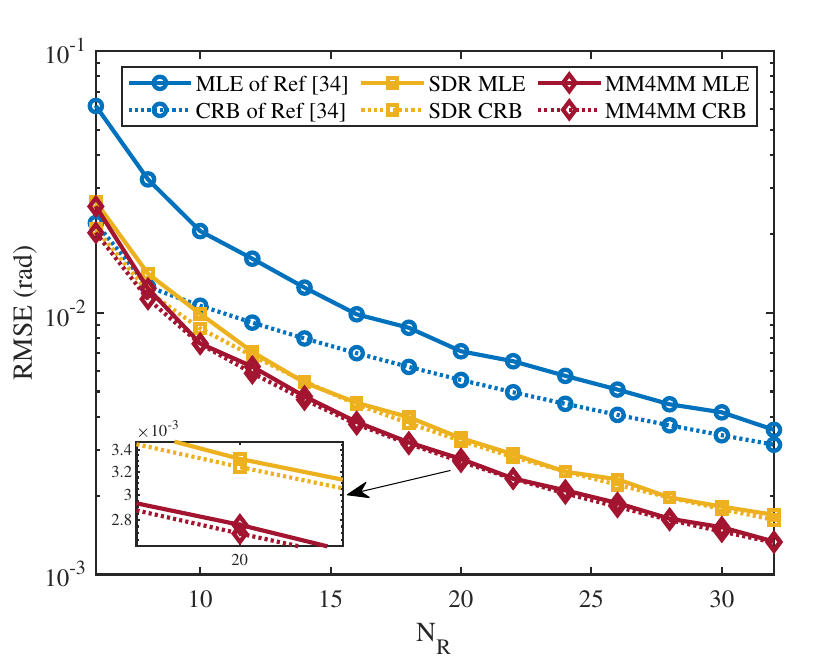}
	\caption{RMSEs versus the number of receiver antennas. $P = 2$. $|\alpha_{1}|^{2} =|\alpha_{2}|^{2} = 0$ dB. $K =6$. $\hat{\Gamma}_{k} = 15$ dB ($k=1,2,\cdots,K$).} \label{CRB_RMSE_NR}
\end{figure}

\figurename ~\ref{CRB_RMSE_NR} compares the RMSEs and the CRBs versus the number of receive antennas for the three algorithms, where the parameters setting is the same as in \figurename ~\ref{Beampattern}, except for the varying number of receive antennas. It can be seen that the MM$4$MM algorithm reaches the lowest CRB for various number of receive antennas. Moreover, even in the case of small $N_\textrm{R}$, in which the asymptotic CRB as well as its upper bound might not be accurate (due to the nonzero cross correlations between the receive array steering vectors at different directions), the gap between the RMSE of the proposed method and the associated CRB is small. 

\begin{figure}[!htbp]
	\setlength{\abovecaptionskip}{0.cm}
	\setlength{\abovecaptionskip}{0.cm}
	\setlength{\belowdisplayskip}{0pt}
	\centering
	\includegraphics[width= 0.42 \textwidth] {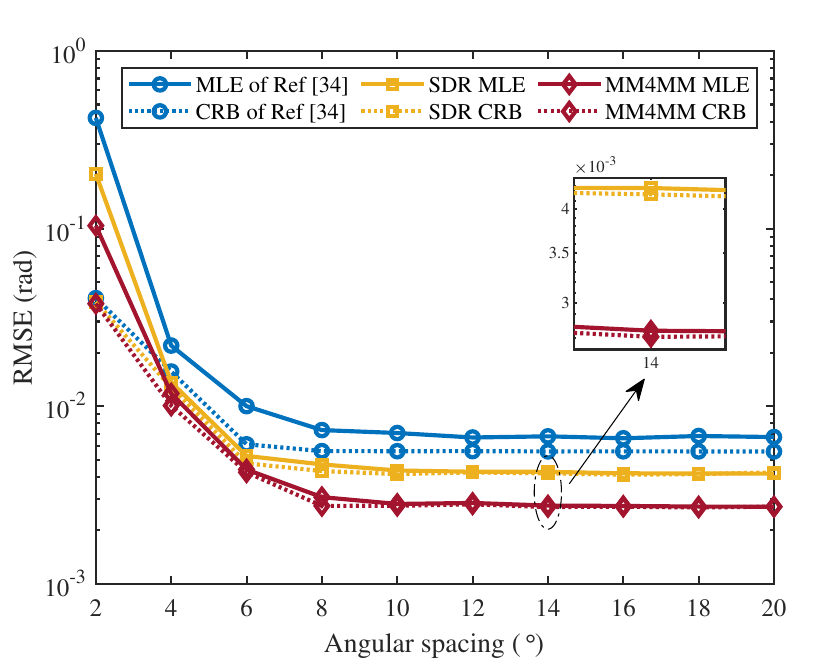}
	\caption{RMSEs versus angular spacing. $P = 2$. $|\alpha_{1}|^{2} =|\alpha_{2}|^{2} = 0$ dB. $K =6$. $\hat{\Gamma}_{k} = 15$ dB ($k=1,2,\cdots,K$).} \label{RMSE_Angular}
\end{figure}

\figurename ~\ref{RMSE_Angular} plots the RMSEs and the CRBs versus the angular spacing for the three algorithms, where the parameter setting is the same as in \figurename ~\ref{Beampattern}, except for the varying angular spacing between the two targets. Note that for a small angular spacing,  the asymptotic CRB as well as its upper bound could be imprecise. However, we can see that for all the angular spacings under consideration, the MM$4$MM algorithm still reaches the lowest RMSE and CRB.

\begin{figure}[!htbp]
	\setlength{\abovecaptionskip}{0.cm}
	\setlength{\abovecaptionskip}{0.cm}
	\setlength{\belowdisplayskip}{0pt}
	\centering
	\includegraphics[width= 0.42 \textwidth] {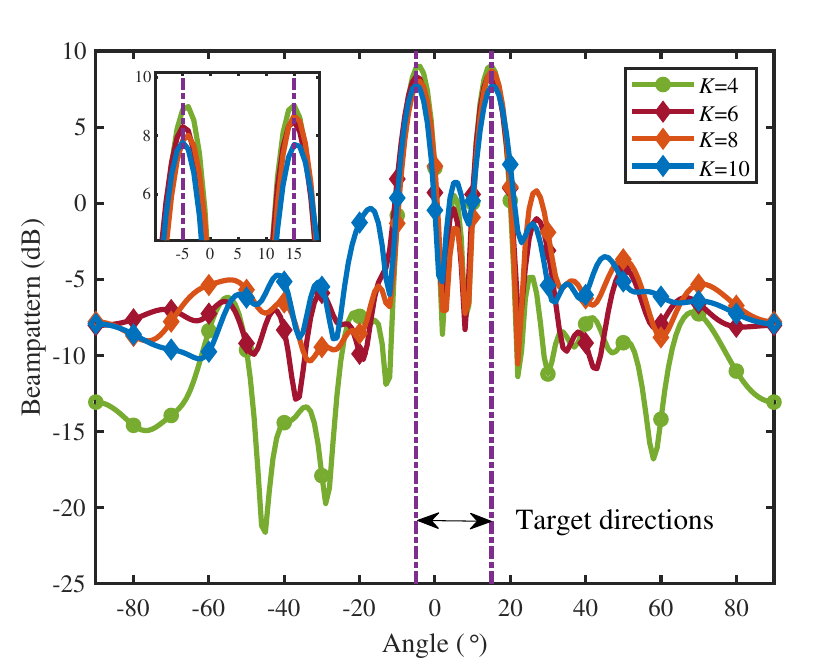}
	\caption{Transmit beampatterns for different number of communication users. $P = 2$. $|\alpha_{1}|^{2} =|\alpha_{2}|^{2} = 0$ dB. $\hat{\Gamma}_{k} = 15$ dB ($k=1,2,\cdots,K$).} \label{Beampattern2}
\end{figure}

Now, we analyze the impact of the number of communication users (i.e., $K$) and the communication SINR threshold (i.e., $\hat{\Gamma}_{k}$) on the target angle estimation performance. \figurename ~\ref{Beampattern2} plots the transmit beampattern associated with the beamforming matrix designed by the MM$4$MM algorithm for different number of communication users, where the parameter setting is the same as in \figurename ~\ref{Beampattern}, except for the varying number of communication users. We can see that with the increasing number of communication users, the beampattern responses at the target directions become weaker, while the sidelobes of the beampattern are higher.

\begin{figure}[!htbp]
	\setlength{\abovecaptionskip}{0.cm}
	\setlength{\abovecaptionskip}{0.cm}
	\setlength{\belowdisplayskip}{0pt}
	\centering
	\includegraphics[width= 0.42 \textwidth] {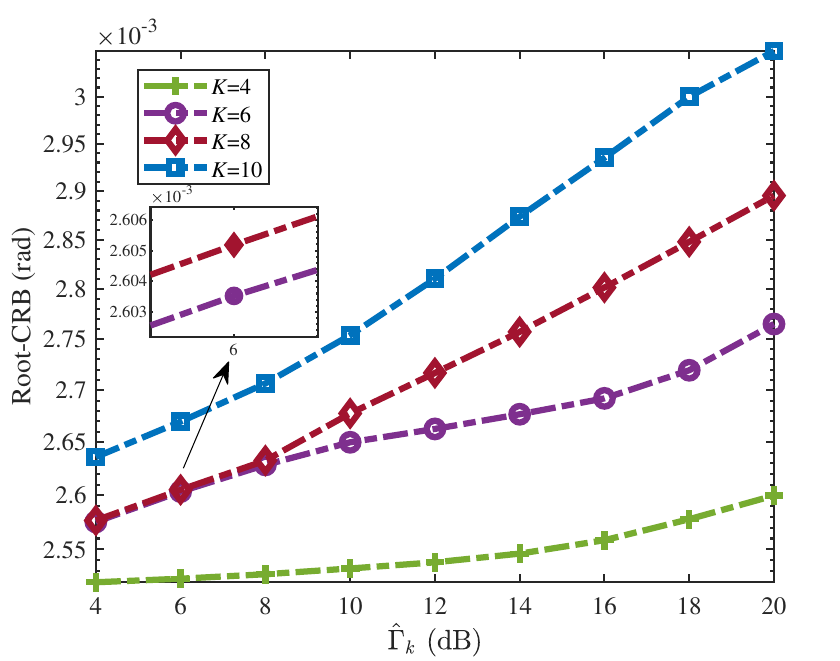}
	\caption{Root-CRB versus the communication SINR threshold for different numbers of users. $P = 2$. $\sigma_{\textrm{R}}^{2} = 0$ dB.}
	\label{Tradeoff}
\end{figure}

\figurename~\ref{Tradeoff} plots the root-CRB versus the communication SINR threshold (i.e., $\hat{\Gamma}_{k}$) and the number of communication users (i.e., $K$) for the beamforming matrix designed by the MM$4$MM algorithm.  The parameter setting is the same as in \figurename~\ref{Beampattern}, except that we change $K$ or $\hat{\Gamma}_{k}$ for each point on these curves. Moreover, $\hat{\Gamma}_{k}, k=1,2,\cdots,K,$ are set to be identical. One can see that the estimation error grows with the number of communication users. Moreover, a more demanding value of the communication SINR also results in a decreased estimation performance.

\begin{figure}[!htbp]
	\centering
{{\includegraphics[width = 0.42\textwidth]{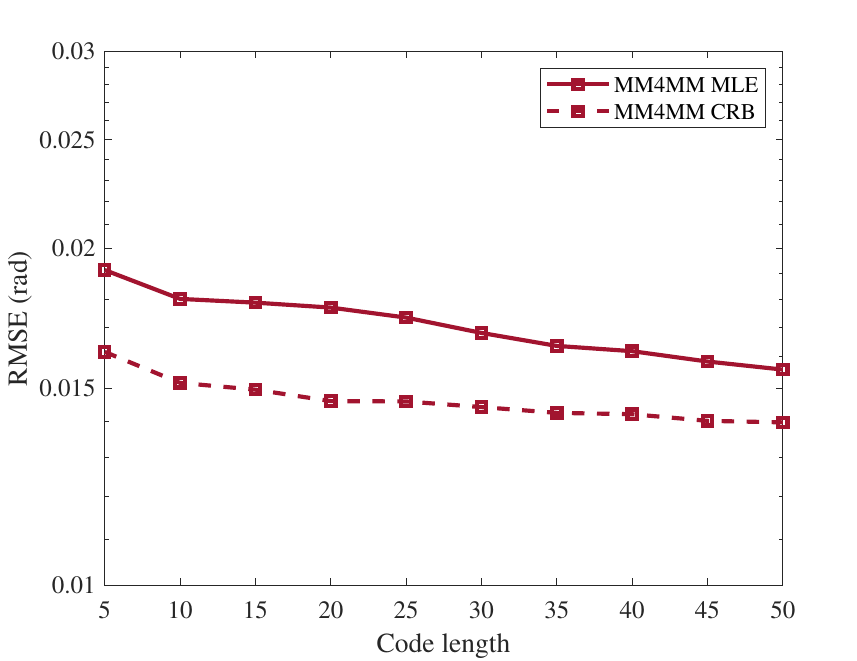}} }
	\caption{RMSE versus code length. $P = 2$. $|\alpha_{1}|^{2} = |\alpha_{2}|^{2} =0$ dB. $K =6$. $\hat{\Gamma}_{k} = 15$ dB ($k=1,2,\cdots,K$).} \label{Fig:CodeLength}
\end{figure}

Finally, \figurename ~\ref{Fig:CodeLength} analyzes the RMSE and the CRB of the beamforming matrix designed by the MM$4$MM algorithm for various code lengths, where the energy of the communication data matrix $\bS$ is fixed to be $K$ (i.e., $\textrm{tr}(\bS\bS^\dagger)=K$), the other parameter setting is the same as in \figurename~\ref{Beampattern}, and $5000$ Monte Carlo trials are conducted to obtain each point on the curves. We can observe that for shorter code lengths, the RMSE and the CRB are slightly higher. Moreover, for code length longer than $25$, the variations of the RMSE and the CRB become insignificant.

\section{Conclusion}\label{Sec:Conclusion}
This paper has considered the design of beamforming matrix for a MIMO DFRC system, whose task is to resolve multiple targets and communicate with multiple users.
To design the dual-function beamforming matrix, we formulated a constrained optimization problem based on minimizing an upper bound on the asymptotic CRB of the target angle estimates. Moreover, an SINR constraint was imposed to guarantee the communication QOS. Two algorithms, named ADMM and MM$4$MM, were developed to solve the non-convex design problem. The results showed that MM$4$MM achieved better performance than the ADMM algorithm but required longer time to converge. In contrast to the competing algorithms, the proposed beamforming matrices formed stronger beampattern responses at the target directions and lower sidelobes. As a result, the proposed designs achieve a lower CRB and thus a better angle estimation performance than the competing design.


\appendices
\section{Proof of Proposition \ref{Prop:1}} \label{Apd:A}
Using the fact that the receive antenna array is a ULA and the result in Appendix G of \cite{STOICA1989MUSIC}, if follows that for $N_{\textrm{R}} \gg 1$, 
\begin{align*}
	\bA_{\textrm{R}}^{\dagger}\bA_{\textrm{R}} \approx N_{\textrm{R}} \bI_{P} , \bA_{\textrm{R}}^{\dagger}\dot{\bA}_{\textrm{R}} \approx \frac{1}{2} \textrm{j} N_{\textrm{R}}^{2} \bI_{P} , \dot{\bA}_{\textrm{R}}^{\dagger}\dot{\bA}_{\textrm{R}} \approx \frac{1}{3} N_{\textrm{R}}^{3} \bI_{P} \textrm{. }
\end{align*}
	
As a result, it can be verified that $\bF_{11}$, $\bF_{12}$, and $\bF_{22}$ are diagonal matrices. Moreover, the $p$th diagonal element of $\bF_{11}$ $(p=1 ,2 ,\cdots ,P)$  is given by
\begin{equation} \label{eq:F11 new}
	\begin{aligned}
		\bF_{11}(p ,p) = &{\frac{1}{3}  N_{\textrm{R}}^{3}|\alpha_{p}|^{2}b_p}  -{\frac{1}{2} \textrm{j}  N_{\textrm{R}}^{2}|\alpha_{p}|^{2} \dot{b}_p} 
		 +{\frac{1}{2} \textrm{j}  N_{\textrm{R}}^{2}|\alpha|_{p}^{2} \dot{b}_p^* }  + N_{\textrm{R}}|\alpha|_{p}^{2} \ddot{b}_p  \\
		=&|\alpha|_{p}^{2}  [\frac{1}{3} N_{\textrm{R}}^{3}b_p  + N_{\textrm{R}}^{2}  \textrm{Im}(\dot{b}_p) + N_{\textrm{R}}\ddot{b}_p ].
	\end{aligned}
\end{equation}
Note that $\bF_{11}(p ,p)$ is real-valued. Similarly, we can verify that
\begin{align}\label{eq:F12 new}
	\bF_{12}(p ,p)
	& =\alpha_p^{*} (-\frac{1}{2} \textrm{j} N_{\textrm{R}}^{2} b_p  + N_{\textrm{R}} \dot{b}_p^*).
\end{align}
and
\begin{align}\label{eq:F22 new}
	\bF_{22}(p , p) =  N_{\textrm{R}} b_p.
\end{align}
By using \eqref{eq:F11 new}, \eqref{eq:F12 new}, and \eqref{eq:F22 new}, one can verify that
\begin{align}
	\bF_{\bm{\omega}} =&\bF_{11}-
	[\begin{matrix}
		\textrm{Re}(\bF_{12}) & -\textrm{Im}(\bF_{12})
	\end{matrix}]  \textrm{BlkDiag}(\bF_{22}^{-1} ;\bF_{22}^{-1} )
[\begin{matrix}
		\textrm{Re}(\bF_{12}) & -\textrm{Im}(\bF_{12})
	\end{matrix}]^\top
	 \nonumber \\
	= &\ \bF_{11} -\bF_{22}^{-1} \bF_{12}\bF_{12}^{*}.
\end{align}
In addition, the $p$th diagonal element of $\bF_{\bm{\omega}}$, denoted $\bF_{\bm{\omega}}(p,p)$, is given by
  \begin{align}\label{eq:F_theta}
	\bF_{\bm{\omega}}(p ,p) ={|\alpha_{p}|^{2}} \left[\frac{1}{12}N_{\textrm{R}}^{3}b_p+N_{\textrm{R}} (\ddot{b}_p -{|\dot{b}_p|^2}{b_p^{-1}})\right].
\end{align}
Thus, \eqref{eq:C_theta} is proved. Moreover, by using the Cauchy-Schwartz inequality, it is easy to verify that
\begin{align}\label{eq:Cauchy}
|\dot{b}_p|^2
= |\dot{\ba}_{p,\textrm{T}}^{\dagger} \bR_{\bX}^{*}\ba_{p,\textrm{T}}|^2 
\le\dot{\ba}_{p,\textrm{T}}^{\dagger} \bR_{\bX}^{*} \dot{\ba}_{p,\textrm{T}} \cdot \ba_{p,\textrm{T}}^{\dagger}\bR_{\bX}^{*}\ba_{p,\textrm{T}}
	= b_p \ddot{b}_p.
\end{align}
Therefore,
\begin{align}\label{eq:F_theta bound}
	\bF_{\bm{\omega}}(p,p) \ge \frac{1}{12}|\alpha_{p}|^2  N_{\textrm{R}}^{3} b_p,
\end{align}
which completes the proof of Proposition \ref{Prop:1}.

\bibliographystyle{IEEEtran}

\bibliography{reference}

\end{document}